\documentclass[conference]{IEEEtran}
\usepackage{mdframed}
\usepackage{tikz}
\usetikzlibrary{tikzmark,fit}
\usepackage{fbox}
\usepackage{float}
\usepackage{xcolor}
\usepackage{framed}
\usepackage{listings}
\usepackage[
  frozencache=true,
  cachedir=minted-cache
]{minted}
\usepackage[normalem]{ulem}
\useunder{\uline}{\ul}{}
\let\labelindent\relax
\usepackage{enumitem}
\usepackage{tikz}
\usepackage{amsmath}
\usepackage{filecontents}
\usepackage[font=small,skip=0pt]{caption}
\usepackage{multirow}
\usepackage{booktabs}
\usepackage{csquotes}
\usepackage{placeins}
\usepackage{xspace}
\usepackage[normalem]{ulem}
\usepackage{url}
\usepackage{subcaption}
\usepackage[most]{tcolorbox}
\usepackage{listings}
\usepackage{breakurl}
\usepackage{hyperref}
\hypersetup{colorlinks,allcolors=black}    
\usepackage{enumitem}
\usepackage{xcolor}
\usepackage{mdframed}
\usepackage{tcolorbox}
\usepackage{tikz}
\usepackage{graphicx}
\usepackage{subcaption}
\usepackage[most]{tcolorbox}
\usepackage{pifont}           

\usepackage{amssymb}         
\usepackage{bbding}          
\usepackage{xcolor} 
\definecolor{lightgreen}{rgb}{0.5, 0.9, 0.2}
\definecolor{codebg}{HTML}{FAFAF7}
\definecolor{guardcloud}{HTML}{E8F1FF}
\definecolor{sigcloud}{HTML}{FFE8D6}
\definecolor{argcloud}{HTML}{E9F7EF}
\definecolor{castcloud}{HTML}{F6E8FF}
\definecolor{patchcloud}{HTML}{FFF4BF}
\definecolor{seqcloud}{HTML}{B7D4FF}
\newcommand{\codehl}[2][sigcloud]{%
  \begingroup\setlength{\fboxsep}{1pt}%
  \colorbox{#1}{\strut\ttfamily\detokenize{#2}}%
  \endgroup}
\colorlet{mygray}{black!30}
\colorlet{mygreen}{green!90!blue}
\colorlet{mymauve}{red!60!blue}
\lstdefinelanguage{diff}{
  morecomment=[f][\color{blue}]{@@},     
  morecomment=[f][\color{red}]-,         
  morecomment=[f][\color{mygreen}]+,       
  morecomment=[f][\color{magenta}]{---}, 
  morecomment=[f][\color{magenta}]{+++},
  morecomment=[f][\color{magenta}]{\#},
}

\DeclareCaptionType{example}[Example][List of Examples]

\newcommand{\system}{\textsc{$\text{SyzHarness}$}\xspace}

\newcommand{\etal}{\emph{et al.}\xspace}
\newcommand{\cut}[1]{}

\usepackage{xstring}
\newcommand{\PP}[1]{
\vspace{2px}
\noindent{\bf \IfEndWith{#1}{.}{#1}{#1.}}
}

\newcommand{\squishlist}{
  \begin{list}{$\bullet$}
    { \setlength{\itemsep}{0pt}      \setlength{\parsep}{3pt}
      \setlength{\topsep}{3pt}       \setlength{\partopsep}{0pt}
      \setlength{\leftmargin}{1.0em} \setlength{\labelwidth}{1em}
      \setlength{\labelsep}{0.5em} } }
\newcommand{\squishend}{
    \end{list}  }

\definecolor{mymauve}{rgb}{0.58,0,0.82}

\begin{document}
\title{SyzHarness: Patch-Based Kernel Bug Reproduction with LLM-Synthesized Fuzzing Harnesses}

\author{
\IEEEauthorblockN{Xingyu Li, Juefei Pu, Haonan Li, Arrdya Srivastav, Kareem Shehada, Srikanth V. Krishnamurthy, Zhiyun Qian}
\IEEEauthorblockA{University of California, Riverside\\
\{xli399,jpu007,hli333,asriv033,ksheh002\}@ucr.edu, \{krish,zhiyunq\}@cs.ucr.edu}
}

\newcommand{\mytodogrey}[1]{\textcolor{mygreen}{\ding{46}~{\sf}~#1}}

\maketitle

\begin{abstract}

Automated kernel vulnerability reproduction is essential for bug triage, patch validation, and regression testing, but still lacks an effective and efficient solution. The core challenge is twofold: a reproducer must first recover the trigger scaffold to reach the vulnerable state {\em and} determine the precise concrete values that actually trigger the bug. Existing directed fuzzing approaches are ineffective in recovering the necessary trigger scaffold, while LLM-only generation is brittle because it struggles with concrete-value discovery and runtime nondeterminism.

We design \system, a framework that combines LLM reasoning with coverage-guided fuzzing for patch-based Linux kernel vulnerability reproduction. Given a patch, \system uses an LLM agent---grounded by code navigation tools ---to synthesize a parameterized fuzzing harness that fixes the prerequisite setup logic while exposing only uncertain, bug-critical input parameters to be mutated by a traditional fuzzer, Syzkaller. \system then translates this harness into a Syzkaller-compatible interface and iteratively refines it using hierarchical reachability feedback.

We evaluate \system on multiple datasets of triggerable real-world Linux kernel vulnerabilities. On 100 KernelCTF cases, \system achieves a 78\% bug reproduction success rate. On the SyzDirect benchmark, \system achieves 73\% bug reproduction success rate, substantially outperforms prior directed greybox fuzzing. On 50 recent, known-triggerable syzbot bugs fixed after March 2026, \system reproduces 40/50 (80\%) using only the fix commits as input. These results show that combining LLM-derived trigger scaffold with fuzzer-driven concrete value search is an effective and efficient approach to automated kernel vulnerability reproduction.

\end{abstract}

\section{Introduction}
\label{sec:intro}

The Linux kernel is one of the most important and security-critical software systems in modern computing, underpinning cloud infrastructure, servers, embedded devices, and mobile platforms. In this setting, \emph{vulnerability reproduction} is a fundamental capability for bug triage, vulnerability impact assessment, and patch validation. A working reproducer
allows developers and analysts to confirm that a bug is real, understand its triggering conditions, and verify that a patch eliminates the root cause without introducing new failures.

In many real-world settings, however, bug reproduction starts not from a crash report or concrete failing input, but from a \emph{security patch}. A public patch does not necessarily coincide with a public PoC. Linux kernel patches are not generally required to be accompanied by a public PoC. Even Google’s Linux kernel vulnerability reward program, kernelCTF~\cite{kernelctf}, requires participants to publish their exploit within 90 days of submitting the patch commit, explicitly allowing publication to be delayed within this window. Consequently, a patch may already be available while the corresponding public trigger is not. This gap matters in practice: upstream maintainers and downstream vendors may need to validate a fix or backport before a usable public reproducer becomes available. Prior work also recognizes missing PoCs as a practical obstacle: SyzDirect notes that Linux Kernel Bugzilla receives many reports without PoCs~\cite{tan2023syzdirect}, while KernJC treats workable PoC availability as a prerequisite for practical kernel-vulnerability reproduction~\cite{ruan2024kernjc}. Our patch-stream study further provides direct evidence of this setting: we collected 64 Linux kernel patches in 2026 for which no public PoC or associated CVE was available at collection time. \system targets this setting: \emph{given a patch but no usable trigger, automatically recover a reproducer for patch validation.}

Patch-based bug reproduction is fundamentally a two-part problem. First, one must recover the \emph{trigger scaffold} needed to reach the vulnerable state: the relevant syscall sequence, protocol steps, object relationships, and state transitions. Second, one must discover the \emph{concrete trigger conditions} that actually manifest the bug, such as precise argument values, boundary cases, timing-sensitive interleavings, and other runtime-dependent effects. The difficulty is that a patch usually exposes the vulnerable region, but reveals only limited information about these hidden prerequisites and trigger conditions.

\noindent\textbf{Limitations of alternative approaches.}
Directed Greybox Fuzzing (DGF), including SyzDirect~\cite{tan2023syzdirect}, is closely related: it steers fuzzing toward vulnerable code (before patching) and further leverages syscall dependencies and argument constraints. However, reaching the patched region is not equivalent to constructing the semantic state required to trigger the vulnerability. Many kernel bugs require a precise syscall sequence, object relationships, protocol state, or resource lifecycle before the patched code becomes vulnerable.

\system therefore differs in what it removes from the fuzzing search space: instead of only guiding executions toward the target, it uses the LLM to synthesize and \emph{fix} a patch-specific trigger scaffold, leaving Syzkaller to search only the remaining uncertain concrete values.

Large Language Models (LLMs) offer a complementary capability. Trained on large corpora containing kernel source code, documentation, and developer discussions, they can often recover information that conventional fuzzers do not explicitly model. When given a patch together with relevant surrounding code, an LLM can often infer the missing \emph{trigger scaffold}
that describes the key subset of requirements needed for triggering the bug: which operations are required, in what order, which subsystems are involved, and which arguments must take specific flags, constants, or structural forms. This makes LLMs naturally attractive for patch-based bug reproduction, because they can recover the setup logic that determines whether execution ever enters the vulnerable state.

However, LLM-only bug reproduction remains brittle. Many vulnerabilities depend not just on recovering the right high-level logic, but also on finding the exact concrete values and runtime conditions that satisfy the final trigger. These may include rare thresholds, unusually large buffer sizes, specific field combinations, allocator-sensitive states, or timing windows. An LLM-generated reproducer may therefore capture a plausible triggering strategy while still failing to realize the precise conditions required to produce the crash. In short, LLMs are comparatively strong at recovering \emph{trigger scaffold}, but weak at searching the remaining \emph{concrete value space}.

These observations suggest a principled hybrid design. The LLM should be used to recover the trigger scaffold that determines \emph{how} to reach the vulnerable state, while the fuzzer should be used to explore the remaining concrete trigger values efficiently. The challenge is how to combine them without losing the strengths of either: the LLM must constrain exploration to a patch-relevant search space, but the fuzzer must still be able to mutate inputs at high throughput.

\noindent\textbf{Our Approach.}
We present \textbf{\system}, a framework that combines LLM reasoning with coverage-guided fuzzing for patch-based Linux kernel vulnerability reproduction. Given a security patch, \system uses an LLM agent to infer the \emph{trigger scaffold} necessary for triggering the bug, including the required setup logic, relevant syscall sequence, and state construction. Rather than asking the LLM to generate a complete reproducer end-to-end, \system turns this structure into a \emph{parameterized harness}: it fixes the high-level setup needed to reach the vulnerable state, while exposing only a small set of uncertain, bug-critical inputs as parameters for automated exploration. The fuzzer then searches this refined space by mutating only the exposed parameters, which is well-suited to discovering precise trigger values and navigating runtime nondeterminism. When bug reproduction fails, \system iteratively refines the harness using reachability feedback, focusing subsequent attempts on missing prerequisites rather than spending the full fuzzing budget on an inadequate setup.

\noindent\textbf{Results.}
We evaluate \system on two datasets of triggerable real-world Linux kernel vulnerabilities. On 100 KernelCTF cases, \system achieves a 78\% bug reproduction success rate with a mean time-to-reproduction of 2.9 hours.
On the benchmark used by SyzDirect~\cite{tan2023syzdirect}, \system achieves a 73\% bug reproduction success rate, significantly outperforming SyzDirect.
On 50 recent syzbot UAF/OOB bugs fixed after March 2026, \system successfully reproduces 80\%.
These results suggest that \system benefits from explicitly separating prerequisite-structure recovery from concrete trigger search, and from assigning these two tasks to LLM reasoning and fuzzing, respectively.

\noindent\textbf{Contributions.}
Our contributions are summarized as follows:
\squishlist
\item \textbf{A new perspective on patch-based kernel bug reproduction. }
We formulate patch-based vulnerability reproduction as a two-level problem: recovering the \emph{trigger scaffold} needed to reach the vulnerable state, and discovering the \emph{concrete trigger values} needed to manifest the bug. This formulation clarifies why fuzzing alone and LLM-only generation fail in complementary ways.

\item \textbf{A principled design combining LLM reasoning and fuzzing.}
We present \system, a hybrid framework that uses an LLM agent to recover patch-relevant setup logic and synthesizes a \emph{parameterized harness} that fixes this structure while exposing only uncertain, bug-critical inputs to fuzzing. This design steers exploration away from unconstrained syscall-program search and toward a patch-relevant search space.

\item \textbf{An end-to-end pipeline for practical bug reproduction.}
We develop a practical pipeline that grounds the LLM in patch and code context, translates the synthesized harness into a fuzzable interface, and iteratively refines it using reachability feedback when initial attempts fail.

\item \textbf{Strong results on real-world vulnerabilities.}
We evaluate \system on KernelCTF, the SyzDirect benchmark, and 50 recent known-triggerable syzbot bugs. \system achieves a 78\% success rate on KernelCTF, substantially outperforms SyzDirect in the controlled comparison, and reproduces 80\% of recent syzbot bugs using only their fix commits.
\squishend
\section{Background}
\label{sec:background}

\subsection{Syzkaller and pseudo-syscalls}

\begin{figure}[t]
    \centering
    \begin{minipage}[t]{0.98\linewidth}
\begin{minted}[
    fontsize=\scriptsize,
    breaklines,
    linenos,
    numbersep=3pt,
    frame=single,
    framesep=1.5mm,
    bgcolor=codebg,
    tabsize=1,
    escapeinside=@@
]{c}
#if @\codehl[guardcloud]{SYZ_EXECUTOR || __NR_syz_open_dev}@

static long @\codehl[sigcloud]{syz_open_dev}@(long a0, long a1, long a2)
{
 if (a0 == 0xc || a0 == 0xb) {
  char buf[128];
  sprintf(buf, "/dev/%s/%d:%d", a0 == 0xc ? "char" : "block", @\codehl[castcloud]{(uint8)a1}@, @\codehl[castcloud]{(uint8)a2}@);
  return open(buf, O_RDWR, 0);
 } else {
  unsigned long nb = a1;
  char buf[1024];
  char* hash;
  strncpy(buf, @\codehl[castcloud]{(char*)a0}@, sizeof(buf) - 1);
  buf[sizeof(buf) - 1] = 0;
  while ((hash = strchr(buf, '#'))) {
   *hash = '0' + (char)(nb % 10);
   nb /= 10;
  }
  return open(buf, a2, 0);
 }
}
#endif
\end{minted}
    \end{minipage}
    \caption{A pseudo-syscall implementation example. The highlights mark
    the guard, function name, and explicit casts.}
    \label{fig:pseudosyscallexample}
    \vspace{-.3in}
\end{figure}

Syzkaller stands as the de facto standard for coverage-guided kernel fuzzing,
having successfully identified over 7,000 vulnerabilities in the Linux
kernel~\cite{syzbot}.
A major reason for its effectiveness is its use of manually curated declarative
\emph{syscall descriptions}, written in \emph{Syzlang}, together with its
mutation-based fuzzing algorithm. Since system calls form the primary interface
between user space and the kernel, Syzkaller must know not only which syscalls
are available, but also how to construct their inputs.

Beyond ordinary syscalls (e.g., open, read), Syzkaller also supports
\emph{pseudo-syscalls}. ~\autoref{fig:pseudosyscallexample} is an example for a pseudo-syscall (more details are introduced later). A native syscall models an existing kernel entry point;
in contrast, a pseudo-syscall is a user-space abstraction implemented inside
Syzkaller's executor and exposed to the fuzzer as if it were an ordinary
syscall. This mechanism allows Syzkaller to package multi-step setup logic or
environment-specific operations that are difficult to express through standard
system call sequences alone, into a single fuzzable interface. Key
functionalities provided by pseudo-syscalls include: (1) {\em Abstraction of Device
Interaction}: facilitating operations on device nodes with non-standard naming
conventions (e.g., syz\_open\_dev);
(2)
{\em Protocol-Specific Injection}: handling the construction and injection of raw
network packets (e.g., syz\_emit\_ethernet); and (3) {\em Complex State Setup}: managing
resource initialization sequences required by subsystems like io\_uring
(e.g., syz\_io\_uring\_setup). By abstracting platform-specific details and state
dependencies, pseudo-syscalls significantly enhance the fuzzer's ability to
reach deep kernel states.

Syzkaller's pseudo-syscall interface is extensible. Developers can define pseudo-syscalls by adding custom descriptions that conform to the interface conventions. This can substantially extend Syzkaller’s capabilities beyond the few use cases in Syzkaller. For example, they can be used to encapsulate a useful sequence of syscalls to exercise a particular functionality in the kernel. However, despite the potential, crafting pseudo-syscalls is predominantly a manual process. It requires developers to possess not only intimate knowledge of the target subsystem (to construct the logic) but also familiarity with Syzkaller’s strict syntax constraints and description language (Syzlang). Consequently, pseudo-syscalls are limited in practice to general-purpose abstractions (e.g., network setup) rather than vulnerability-specific reproduction scenarios.

\section{Design Rationale}

Our methodology is driven by the following challenges and insights related to the limitations of current automated vulnerability reproduction methods.

A bug reproduction artifact must recover two core ingredients: the syscall sequences
that lead to the function that carries the bug and the values of syscall
arguments that cause the bug to be triggered. Determining these two attributes
constitutes the major challenge in bug reproduction. In other words, a
patch-based reproducer must recover two complementary ingredients: (i) a
\emph{syscall sequence} that drives the kernel into the vulnerable state (i.e.,
the correct syscall ordering and prerequisite setup), and (ii) \emph{concrete
argument values} that satisfy the bug-triggering conditions.

\subsection{Motivating example}

\begin{figure}[t]
    \centering
    \begin{minipage}[t]{0.98\linewidth}
\begin{minted}[
    fontsize=\scriptsize,
    breaklines,
    frame=single,
    framesep=1.5mm,
    bgcolor=codebg,
    tabsize=2,
    escapeinside=||
]{diff}
scsi: scsi_ioctl: Validate command size

Need to make sure the command size is valid before 
copying the command from user space.

--- a/drivers/scsi/scsi_ioctl.c
+++ b/drivers/scsi/scsi_ioctl.c
@@ -347,6 +347,8 @@ static int scsi_fill_sghdr_rq(struct 
scsi_device *sdev, struct request *rq,
 {
 	struct scsi_request *req = scsi_req(rq);
 
+	if (hdr->cmd_len < 6)
+		return -EMSGSIZE;
 	if (copy_from_user(req->cmd, hdr->cmdp, 
        hdr->cmd_len))
 		return -EFAULT;
 	if (!scsi_cmd_allowed(req->cmd, mode))
\end{minted}
    \end{minipage}
    \caption{The patch adds a lower bound on \texttt{SG\_IO} command length before the kernel copies and later interprets a userspace SCSI command.}
    \label{fig:motivateexample}
    \vspace{-.3in}
\end{figure}

\autoref{fig:motivateexample} illustrates this decomposition using a compact
SCSI \texttt{SG\_IO} bug. The patch adds an early termination when
\texttt{hdr->cmd\_len < 6} in \texttt{scsi\_fill\_sghdr\_rq()}. Before the patch, the kernel can copy any \texttt{hdr->cmd\_len} bytes
from user space into the request and later allow the SCSI request path to
interpret that command state. The vulnerable kernel crashes in \texttt{scsi\_queue\_rq()}, along the path
\texttt{sg\_io()} $\rightarrow$ \texttt{scsi\_fill\_sghdr\_rq()} $\rightarrow$
\texttt{scsi\_setup\_scsi\_cmnd()}
$\rightarrow$
\texttt{scsi\_command\_size()}.

Reproducing this bug requires more than noticing the new inequality. First, the
reproducer must construct the correct structural path: it must open a usable
SCSI endpoint, populate a valid \texttt{sg\_io\_hdr}, and issue
\texttt{ioctl(..., SG\_IO, ...)} so that the execution reaches
\texttt{scsi\_fill\_sghdr\_rq()}. Second, it must satisfy a sparse concrete tuple:
the selected device must correspond to the vulnerable block \texttt{SG\_IO}
path; the opcode must survive \texttt{scsi\_cmd\_allowed()} and the transfer
metadata must pass surrounding \texttt{sg\_io} validation. Most importantly,
although the patch suggests that the bug lies in the range
\texttt{cmd\_len < 6}, not all values in that range are equally effective. In
the vulnerable path, the observed crash is triggered by \texttt{cmd\_len = 0}, but
not by other values such as \texttt{5}. The reason is that
the subsequent dangerous path
is executed only when
\texttt{cmd\_len == 0}; with \texttt{cmd\_len = 5}, the kernel instead
continues down a normal error path instead. \autoref{fig:motivateexample} is therefore a
concrete example of a bug whose patch exposes a relevant field, but not the
exact desired value.

\subsection{Why fuzzing (alone) struggles: missing trigger scaffold}

As the Linux kernel continues to grow in size and complexity, modern Linux kernel vulnerabilities are increasingly hidden behind \emph{precondition barriers}: deep state dependencies, subsystem-specific protocols, and implicit preconditions that are not evident from a patch diff alone.

Coverage-guided fuzzing  repeatedly generates syscall programs,
executes them, and retains those inputs that exercise new code. At a high
level, this is a brute-force exploration strategy over a very large space of
possible syscall sequences and argument values. This search space is already
challenging even when the fuzzer has syscall descriptions for all relevant syscalls.
For patch-based bug reproduction, the difficulty is greater: the fuzzer is rewarded
for executing new code, not for satisfying the specific conditions required
to trigger the target bug.

This makes the search highly inefficient. Many kernel subsystems require a
particular setup routine, a specific ordering of calls, and valid object
lifetimes and relationships. Mutation-based fuzzers modify existing programs
incrementally, for example, by changing argument values and/or inserting or deleting
individual calls. Such changes are
effective when the target can be reached via gradual refinement, but they
are poorly suited to reconstructing an entirely different protocol sequence or
state-construction procedure from scratch. As a result, without knowledge of the
required trigger scaffold, the fuzzer spends much of its budget exploring
executions that increase code coverage but remain irrelevant to bug triggering.

Directed greybox fuzzing (DGF) partially mitigates this by prioritizing executions that appear \emph{close} to the patched site (e.g., in terms of control-flow distance). However, distance is not the same as \emph{reachability} and reachability is not the same as \emph{triggerability}. A control-flow path may exist on the CFG and yet remain effectively blocked by data-flow constraints and semantic checks (e.g., “the object must be initialized by a specific handshake,” “a flag must be set via a separate configuration call,” or “a resource must be in a particular internal state”).
Consequently, when a fuzzer encounters a guard such as \texttt{if(state == READY)} (note that it is not \texttt{if(input == CONST)}), it often effectively engages in an unguided search, as a fuzzer is not aware of the ``state machine'' of the target module. This can result in a very large number of trials before the correct state construction is discovered. This phenomenon has been documented as the well-known “dependency challenge” in kernel fuzzing~\cite{hao2022demystifying}.

The SCSI case in \autoref{fig:motivateexample} is representative. A fuzzer that
starts from ordinary syscall programs must simultaneously discover a usable SCSI
endpoint, the \texttt{SG\_IO} ioctl interface, a command
opcode that is accepted by \texttt{scsi\_cmd\_allowed}, and compatible transfer
metadata. Even if it reaches code near \texttt{drivers/scsi/scsi\_ioctl.c},
most mutations are rejected before the request reaches the vulnerable queueing
path. In other words, coverage near the patched file does not imply that the
fuzzer has discovered the trigger scaffold, let alone the concrete triggering
tuple.

\subsection{Why LLMs (alone) are insufficient: inferring precise concrete values}

Despite their strong semantic reasoning capabilities, LLMs are not reliable at readily producing the \emph{exact} concrete inputs required to trigger many kernel bugs.

A major challenge lies in discovering the precise attribute values needed to trigger the bug.
Triggering conditions often depend on boundary values, size thresholds, or carefully chosen field combinations. Even when an LLM correctly infers the form of the constraint (e.g., “this length must exceed a limit,” “this value must overflow,” or “this buffer must be unusually large”), the exact value is often not recoverable from the patch and code context alone. This limitation is not merely hypothetical: in our evaluation, we observe cases where an LLM-only approach fails, whereas our proposed method succeeds by leaving such uncertain choices to systematic exploration. The  case in \autoref{fig:motivateexample} is one such example.

For the SCSI case in \autoref{fig:motivateexample}, the LLM-only approach
correctly infers the high-level trigger scaffold and identifies
\texttt{cmd\_len} as the critical field. However, it instantiates that field
with the non-triggering value \texttt{5}. This is insufficient because on the vulnerable block \texttt{SG\_IO} path, \texttt{cmd\_len = 5} does
not enter the dangerous zero-length fallback in
\texttt{scsi\_setup\_scsi\_cmnd}; the observed crash is triggered by
\texttt{cmd\_len = 0}. The failure is therefore not a lack of structural
understanding; it is the inability of an LLM-only approach to perform
systematic search over a sparse concrete space consisting of command
length, device class, opcode, and transfer metadata.

Even when the correct values could in principle be found through repeated LLM attempts, doing so is costly because the search space is large and sparse.
Consequently, directly asking an LLM to generate a complete reproducer often
fails not because it misses the overall triggering logic, but because it does
not supply the precise concrete values required to cross the final trigger
condition. Indeed, we find that an LLM-only solution fails to trigger this bug in our experiments reported in \S\ref{sec:eval}.

\vspace{-.2in}
\subsection{Fuzzing and LLMs are complementary}
\label{sec:comp}

Fuzzing lacks semantic guidance, but we make the observation that an LLM can help: given the patch context (and surrounding code when needed), an LLM can analyze likely trigger conditions, hypothesize the prerequisite sequence of operations, identify which syscalls are likely involved, and recognize that certain arguments must come from specific enumerations, flags, or header-defined constants.
In the SCSI case (\autoref{fig:motivateexample}), this semantic step recovers the correct trigger scaffold: open a usable SCSI endpoint, construct a valid \texttt{sg\_io\_hdr}, and issue \texttt{ioctl(..., SG\_IO, ...)} and thus, the execution reaches the vulnerable request path. As prior work has shown, inferring such dependencies among syscalls and subsystem-specific interfaces is inherently difficult for traditional methods~\cite{hao2022demystifying}.
Such guidance can help approach the target region and assist in the construction of the required vulnerable state. Moreover, LLM inference need not rely solely on parametric knowledge; we can supply targeted external references via retrieval-augmented prompting when a more complex setup is needed to prepare the vulnerability state.

LLMs fall short in inferring the precise and concrete values for bug triggering,
but fuzzing can help. First, fuzzers can test large numbers of input variants at
high throughput (e.g., Syzkaller can try thousands of inputs in one minute using a single core on a modern server)
making them
well-suited to discovering boundary values, rare combinations, and thresholds
that are difficult to guess. Second, by repeatedly executing slightly different
inputs, fuzzing naturally samples runtime effects (e.g., different allocation
patterns and interleavings), increasing the probability of encountering the
nondeterministic conditions required to produce a crash. In short, once the
\emph{trigger scaffold} of a bug is approximately correct, fuzzing is
an efficient mechanism for exploring the remaining degrees of freedom.

Based on the above, one can conclude that LLMs and fuzzing are complementary for triggering a vulnerability: LLMs are adept at recovering the \emph{trigger scaffold} of a trigger (syscall sequencing, state transitions, and protocol adherence), while fuzzing is effective at discovering \emph{concrete values} and exploring \emph{runtime nondeterminism}.

Our design goal is therefore to combine them without losing the strengths of either:  preserve fuzzer-driven mutation for concrete search while constraining exploration to a trigger scaffold that is relevant to the target bug. Returning to the SCSI example in \autoref{fig:motivateexample}, an ideal method would fix the \texttt{open} $\rightarrow$ \texttt{SG\_IO} scaffold and the construction of the key \texttt{sg\_io\_hdr} argument, while leaving \texttt{cmd\_len}, opcode, transfer metadata, timeout, flags, and device selection as variables for the fuzzer to explore.

\section{\system design and implementation}
\label{sec:method}

\subsection{Overview}
\label{sec:method-overview}

\begin{figure*}
\centering
  \includegraphics[width=0.8\textwidth]{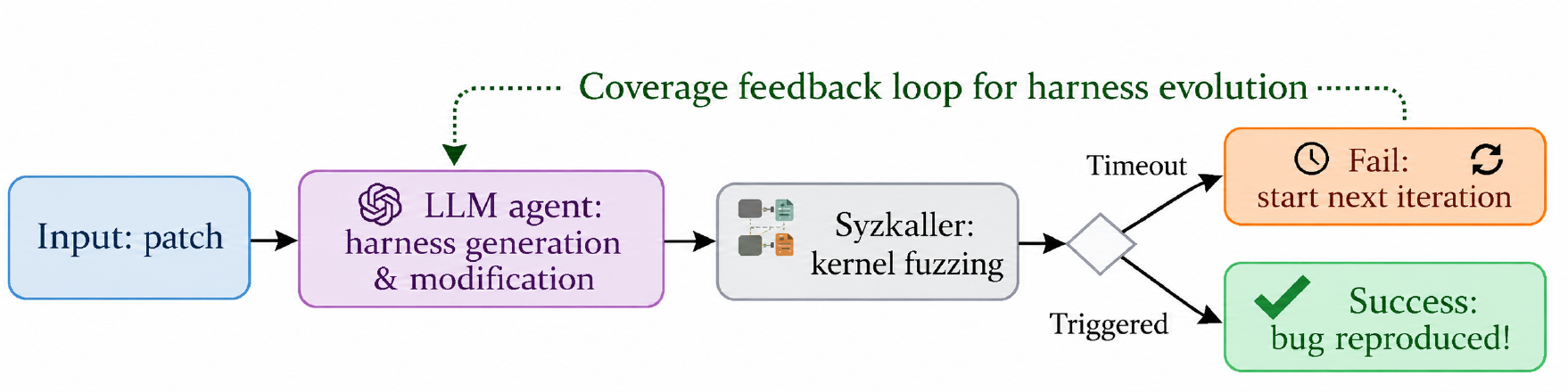}
  \caption{The simplified pipeline of \system }
  \label{fig:pipeline_simplify}
\vspace{-.3in}
\end{figure*}

Given a security patch (commit message + diff), \system aims to trigger the vulnerability fixed by that patch on the corresponding pre-patch kernel version. As discussed in Section~\ref{sec:comp}, LLMs and fuzzing are naturally complementary: LLMs are effective at inferring prerequisite structure for bug triggering, whereas fuzzing is effective at discovering precise concrete values and navigating runtime nondeterminism. An intuitive direction is therefore to combine the two. However, turning this intuition into a practical bug reproduction pipeline introduces several challenges.

\noindent\textbf{Design challenges.}
The first challenge is how to \emph{bridge the representation gap} between LLM reasoning and fuzzing. Traditional Syzkaller-based workflows operate
on syscall descriptions and mutate syscall programs. One possible approach is to let the LLM generate a seed program directly, such as SyzGPT~\cite{zhang2025unlocking}. However, this is ineffective for targeted bug reproduction: once fuzzing begins, coverage-guided mutation can quickly drift by appending unrelated syscalls that increase coverage but do not help trigger the target vulnerability. As a result, the search space remains effectively unconstrained. Moreover, this approach relies on existing syscall descriptions, which can be incomplete~\cite{hao2022demystifying}; missing knowledge typically pertains to syscall relationships, argument types, and ranges.

The second challenge is how to preserve the \emph{targeted structure} inferred by the LLM while still allowing the fuzzer to explore uncertain parts efficiently. For patch-based bug reproduction, the critical difficulty is often not merely reaching the right subsystem, but executing the correct prerequisite sequence with the right state construction. These structural constraints are usually not captured well by syscall descriptions alone, and are difficult to enforce if the LLM output is translated directly into ordinary fuzzing inputs.

The third challenge is \emph{practical integration}. Even if the LLM can synthesize a useful structured reproducer, that artifact must still be transformed into a form that a state-of-the-art kernel fuzzer such as Syzkaller, can schedule, execute, and mutate efficiently. This requires bridging not only a conceptual gap, but also an interface gap between the generated harness code and the fuzzer’s native execution model.

\noindent\textbf{Overview of our solution.}
The high-level workflow of \system is illustrated in \autoref{fig:pipeline_simplify}. Our solution is to use the LLM to synthesize a \emph{parameterized fuzzing harness} that (i) fixes the syscall sequence and state construction required to reach the vulnerable context, and (ii) exposes only uncertain, bug-critical ``knobs'' (e.g., sizes, flags, buffer contents, and selected fields) as explicit parameters for mutation. The fuzzer then takes this harness as the execution scaffold and mutates only the exposed parameters. This design allows the fuzzer to efficiently search the remaining concrete space, while avoiding the combinatorial blow-up caused by semantically irrelevant syscall programs. In effect, \system transforms the search from unconstrained syscall-program exploration into exploration within a constrained, semantically valid subspace.

In the SCSI example from \autoref{fig:motivateexample}, this design yields a
pseudo-syscall with the interface
\texttt{entry(cmd\_len, opcode, dxfer\_len, dxfer\_dir, timeout\_ms, flags, device\_index)}.
The harness fixes the \texttt{open} $\rightarrow$ \texttt{SG\_IO} scaffold and
constructs a valid \texttt{sg\_io\_hdr} internally, while the generated
\texttt{syzlang} restricts \texttt{cmd\_len} to the vulnerable regime and
allows Syzkaller to mutate the remaining command tuple. In our runs, this
combination reaches the vulnerable request path and triggers the observed
general protection fault in \texttt{scsi\_queue\_rq}.

To realize this design, \system incorporates three stages:
\squishlist
  \item \textbf{Harness synthesis from a raw patch.}
  An LLM agent analyzes the patch and surrounding code context to infer likely triggering preconditions; then, it synthesizes a C/C++  \emph{fuzzing harness} that fixes the required syscall ordering and state construction while \emph{parameterizing} uncertain, bug-critical values (e.g., sizes, flags, buffer contents).

  \item \textbf{Bridging the harness to Syzkaller.}
 To make the harness executable under fuzzer-driven mutation, we automatically translate the harness into (i) a Syzkaller \emph{pseudo-syscall} implementation embedded in the executor and (ii) a corresponding \texttt{syzlang} \emph{syscall description} for the harness’s exposed parameters. A lightweight repair agent resolves any build failures introduced during integration.

  \item \textbf{Iterative refinement with reachability feedback.}
If fuzzing does not reproduce a crash within a time budget, we terminate the run, collect hierarchical coverage signals (file $\rightarrow$ function $\rightarrow$ target-site), and feed them back to the LLM to revise the harness for the next iteration. We allocate larger budgets to later iterations and terminate early when the harness fails basic reachability checkpoints.

\squishend

This design separates responsibilities cleanly: the LLM fixes the prerequisite structure that defines a semantically valid search corridor and \emph{prunes exploration space} by locking in state construction, while the fuzzer performs high-throughput  \emph{concrete mutation} over the remaining uncertain and exposed parameters.

\vspace{-.1in}
\subsection{LLM agent for harness synthesis}

\subsubsection{Harness synthesis grounded in patch and code context}
A security patch (commit message and diff) rarely contains sufficient context to derive a correct reproducer
in isolation. Although LLMs encode substantial prior knowledge about Linux kernel APIs and common subsystem patterns, relying solely on parametric knowledge is brittle: patches often reference macros, called functions, and subsystem invariants defined elsewhere, and omitting that context can lead to incorrect assumptions and invalid harnesses.

To improve reliability, \system allows the LLM to retrieve grounded code context beyond the diff itself, including the precise definitions and usage sites of symbols  (e.g., variables, macros, constants, and called functions) referenced by the patch, as well as surrounding call paths and related code regions that govern reachability and triggering. This additional context serves two purposes. First, it resolves ambiguities that cannot be determined from the diff alone, such as the meaning of a flag, the expected state of an object, or the preconditions enforced by a helper function. Second, it anchors the agent's reasoning in the actual kernel implementation, reducing reliance on conjecture and thereby lowering the probability of synthesizing invalid setup sequences.

Accordingly, \system instantiates the LLM as an \emph{agent} equipped with code-navigation tools, rather than as a pure code generator. These tools include symbol definition lookup for functions, macros, and variables; reference tracing to identify where a symbol is used; text-pattern search over the kernel codebase; and targeted source retrieval by file and line range.
We further constrain the agent with a strict \emph{hypothesize--verify--generate} workflow:

\begin{enumerate}
  \item \textbf{Hypothesize:} Derive candidate trigger preconditions from the commit message and modified code paths (e.g., required call sequence, critical flags, or resource lifecycle constraints).
  \item \textbf{Verify:} Query the codebase for referenced symbols, macro meanings, and relevant call paths to confirm or refute the hypothesis (e.g., locate where a flag is consumed, validate which branch guards the vulnerable operation).
  \item \textbf{Generate:} Synthesize harness logic only after the trigger scaffold is grounded in concrete source context.
\end{enumerate}

\subsubsection{External knowledge retrieval for complex syscall usage}
LLMs typically perform well on \emph{common} kernel interaction patterns that appear frequently in public code and discussions—for example, creating an IPv6 socket,
configuring it with standard options, and sending messages via the socket. However, patch-based bug reproduction may require \emph{rare or highly structured} subsystem setups whose correct execution depends on intricate resource lifecycles, non-obvious ordering constraints, or environment-specific conventions. Examples include \texttt{io\_uring} initialization and submission workflows, filesystem mounting sequences, and operations on device nodes with non-standard naming or discovery requirements (e.g., DRI device nodes). Because such procedures are relatively underrepresented (and often very different from general patterns) in typical training corpora, purely parametric LLM reasoning can omit critical steps or propose invalid sequences.

To improve robustness in these cases, \system augments the agent with \emph{curated exemplar implementations} from Syzkaller. Syzkaller maintainers encode hard-won domain knowledge as pseudo-syscalls (as discussed earlier) that implement canonical setup routines for complex subsystems. When \system detects that an existing pseudo-syscall is relevant to the patched code path (e.g., it exercises syscalls in the same subsystem), we retrieve and supply that pseudo-syscall to the agent as external context. The agent can then (i) reuse the pseudo-syscall logic directly as part of the harness, or (ii) mine it for ordering and argument constraints, to  adopt those to satisfy patch-specific trigger conditions. This retrieval mechanism reduces brittle guesswork and strengthens harness correctness for complex, low-frequency kernel state construction.

\subsubsection{Harness generation with explicit parameter exposure}
\label{sec:method-harness}
Given validated preconditions, the agent synthesizes a fuzzing harness that decomposes bug reproduction into:
(i) \emph{fixed triggering logic} (state construction + syscall sequence) and,
(ii) \emph{mutable parameters} that the LLM agent treats as uncertain and leaves intentionally under-specified for fuzzing.

Concretely, the harness includes (as applicable):
\squishlist
  \item \textbf{Environment setup:} required namespaces, module loading, device discovery/opening, and prerequisite configuration.
  \item \textbf{Dependency construction:} creation/initialization of kernel-facing objects and resources needed by the target subsystem.
  \item \textbf{Sequence definition:} a concrete ordering of syscalls to drive the kernel into the vulnerable state.
\squishend

Crucially, the harness only \emph{exposes uncertain, bug-critical knobs} as entry-function parameters (e.g., sizes, flags, buffer contents, selected scalar fields). This parameterization is central: it prevents the fuzzer from wasting effort on semantically invalid programs, while still allowing efficient exploration of the remaining concrete space.

The SCSI case in \autoref{fig:motivateexample} is a direct example of exposing
only uncertainty. The harness constructs \texttt{struct sg\_io\_hdr}
internally and fixes the \texttt{SG\_IO} invocation pattern, but exposes only
those fields whose exact values are uncertain: command length, opcode, transfer
direction and length, timeout, flags, and device index. This lets Syzkaller
search the small, environment-dependent triggering tuple without wasting effort
on unrelated file operations or malformed ioctl layouts.

\subsubsection{Harness verification to reduce hallucinations}
To reduce syntax and environmental errors, the agent is given access to compilation checks in the target (or a faithfully reconstructed) kernel build environment. The harness is compiled automatically  before integration. This step catches common failure modes early (missing headers, wrong types, incorrect API usage), improving the stability of downstream fuzzing.

\subsection{Integrating the harness into Syzkaller}
\label{sec:method-syzkaller}

Syzkaller fuzzes programs described in \texttt{syzlang} and executes them via a dedicated executor. To make a synthesized harness fuzzable, \system converts it into a Syzkaller \emph{pseudo-syscall} plus a corresponding syscall description.

\subsubsection{Syscall description synthesis in the same semantic context}
After harness generation, the agent produces a syscall description for the harness entry function so that Syzkaller can mutate the exposed parameters. Two design choices make this tractable and robust:

\squishlist
  \item \textbf{Session consistency:} harness generation and description generation occur in the same agent session and so, argument meaning and typing decisions remain aligned (avoiding mismatches common in multi-stage pipelines).
  \item \textbf{Interface simplification:} because the pseudo-syscall is synthesized from scratch by the agent, its interface can be deliberately designed to expose only \emph{scalar} parameters (e.g., integers, flags, and byte arrays), rather than deeply nested or complex structures. This keeps the syscall description compact and avoids the need for heavyweight specification inference. Any required nested structures can then be constructed internally within the fuzzing harness from these exposed scalar inputs.
\squishend

When viable, the agent retrieves and reuses existing Syzkaller type definitions (e.g., standard flag sets) in lieu of re-defining them, improving compatibility and reducing syntax errors.

\subsubsection{Pseudo-syscall transformation}

Syzkaller pseudo-syscalls must follow strict executor conventions. \system
applies a deterministic rewrite from the standalone harness function into a
pseudo-syscall entry. Specifically, it:

\begin{enumerate}
    \item adapts the function signature to the executor-required form, for example
    \texttt{static long syz\_<name>(volatile long a0, ...)};

    \item inserts explicit type casts from executor argument-passing conventions
    into the original argument types; and

    \item wraps the implementation with appropriate preprocessor guards, for example
    \texttt{\#if SYZ\_EXECUTOR || defined(\_\_NR\_syz\_name)}.
\end{enumerate}

This transformation is mechanical and repeatable, ensuring that semantic
decisions remain in the harness (LLM-guided) while executor conformance is
handled systematically.

\subsubsection{Compilation repair agent}
\label{sec:method-repair}
Injecting a new pseudo-syscall and its syscall description can still trigger
build failures, primarily because pseudo-syscalls share a common compilation
unit (e.g., \texttt{executor/common\_linux.h}), where newly added code may
conflict with existing includes, types, or definitions. To ensure end-to-end
automation without compromising the fuzzer, \system employs a \emph{lightweight
repair agent} with tightly scoped permissions:

\squishlist
  \item \textbf{Inputs:} compiler error logs and the exact line ranges of the injected code.
  \item \textbf{Scope restriction:} the agent may edit \emph{only} the newly injected pseudo-syscall and description blocks; it is not permitted to modify Syzkaller’s pre-existing code (including maintainer-written pseudo-syscalls).
  \item \textbf{Objective:} resolve compilation errors within at most five repair rounds, stopping early once the executor builds successfully.
\squishend

This constraint preserves the integrity of the underlying fuzzer while enabling fully automated integration.

\subsection{Hierarchical coverage-guided refinement}
\label{sec:method-feedback}

Even encoded with high-level semantics, an initial harness can be incomplete (missing a prerequisite step) or mis-specified (wrong flag ranges, wrong ioctl command, etc.). \system therefore uses coverage-guided feedback not as a direct objective for bug reproduction, but as a \emph{diagnostic signal} for iterative harness repair.

\subsubsection{Isolated fuzzing configuration}
To focus computation on patch-relevant logic, we run Syzkaller with a restricted configuration that enables only the newly crafted pseudo-syscall. This prevents the fuzzer from drifting into unrelated subsystems and ensures that observed coverage changes are attributable to the harness and its exposed parameters.

\subsubsection{Hierarchical reachability signals}
If fuzzing fails to trigger a crash within a per-iteration timeout, we terminate the run and collect hierarchical reachability signals:

\begin{enumerate}
  \item \textbf{File-level reachability ($T_\text{file}$):} whether the patched source file is executed.
  \item \textbf{Function-level reachability ($T_\text{func}$):} whether the patched function(s) are executed.
  \item \textbf{Line-level reachability ($T_\text{line}$):} whether the specific lines modified/removed by the patch are executed.
\end{enumerate}

These signals are returned to the synthesis agent, which revises the harness accordingly (e.g., fix driver/device initialization if $T_\text{file}$ fails; refine command codes/flags if $T_\text{func}$ fails; adjust subtle state constraints if $T_\text{line}$ fails). The syscall description is updated in lockstep with any parameter/interface changes.

\subsubsection{Adaptive time budgeting and early termination}
Harness quality typically improves across iterations as missing prerequisites are identified and corrected. We therefore allocate larger fuzzing budgets to later iterations. Concretely, we run a fixed number of iterations with monotonically increasing timeouts (e.g., 1h, 2h, 4h, 6h, 8h), and stop early on clearly non-viable harnesses.
This schedule is chosen to fit within a practical end-to-end bug reproduction budget (approximately 24 hours), which includes not only fuzzing time but also harness synthesis, pseudo-syscall generation, and compilation/repair. We treat runs that do not reproduce within this budget as failures, consistent with common operational expectations for fuzzing-based methods\cite{tan2023syzdirect,wang2021syzvegas,hao2023syzdescribe}.

To avoid wasting resources, we adopt an early termination policy: if a harness fails to reach $T_\text{file}$ within an initial fraction of its allocated budget (e.g., the first 30\%), we terminate the iteration and trigger immediate harness revision. This is driven by the empirical observation that viable harnesses typically can accomplish the desired maximum coverage quickly (e.g., within 10 minutes); prolonged failures at $T_\text{file}$ strongly indicate a missing prerequisite rather than insufficient mutation time.

\subsubsection{Iterative fuzzing harness}
Each iteration produces a new harness version, regenerates (or updates) its pseudo-syscall and description, recompiles Syzkaller, and starts fuzzing in a new iteration. The agent retains prior iteration context to support monotonic improvement instead of restarting from scratch. The pipeline terminates once the bug is triggered or the time limit is reached.

\begin{figure*}
\centering
  \includegraphics[width=0.8\textwidth]{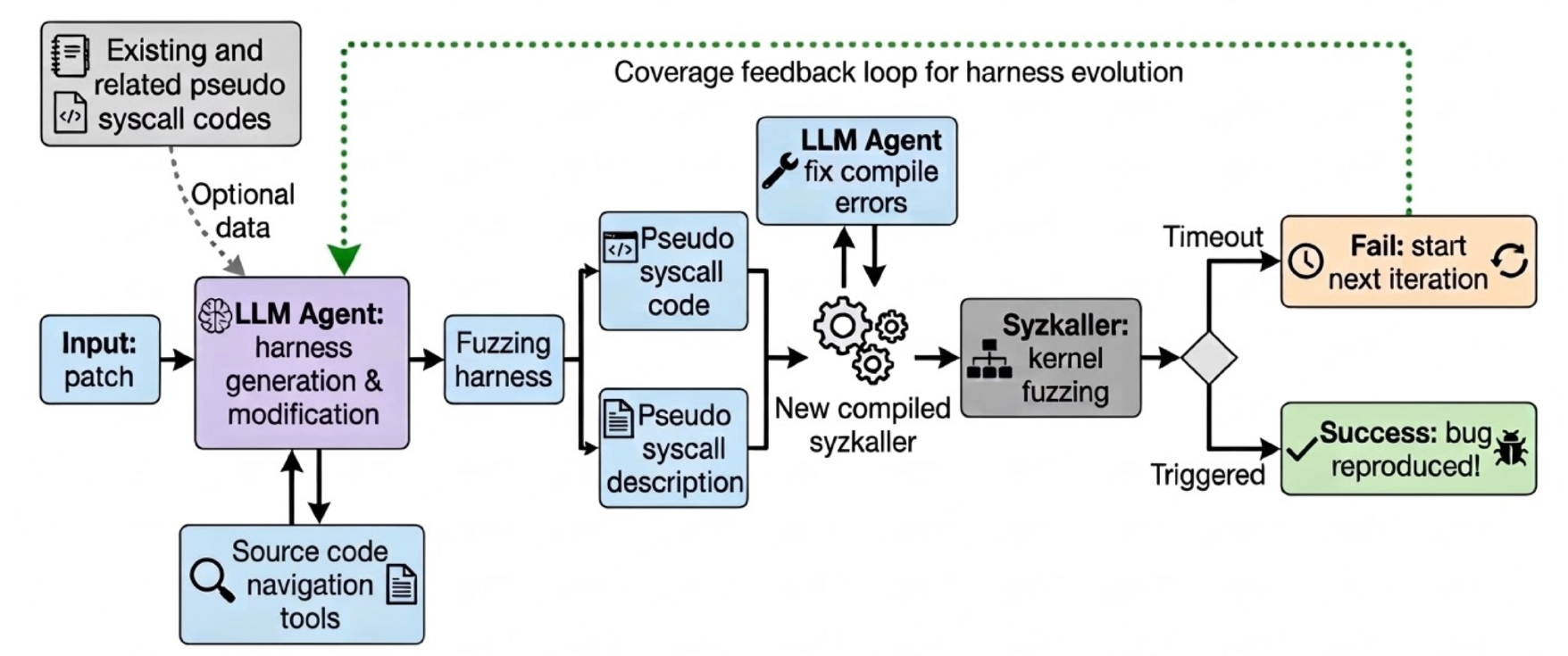}
  \caption{The complete pipeline of \system }
  \label{fig:pipeline_full}
\vspace{-.3in}
\end{figure*}

The overall pipeline is shown in ~\autoref{fig:pipeline_full}.
Starting from an input patch, \system first invokes an LLM agent to analyze the patch and synthesize a fuzzing harness, aided by source-code navigation tools and optionally existing related pseudo-syscall implementations. The generated harness is then translated into two artifacts: pseudo-syscall code and a corresponding syscall description. These artifacts are injected into Syzkaller and compiled into a new fuzzer instance; if compilation fails, a lightweight repair agent iteratively fixes only the injected code until the build succeeds. The resulting Syzkaller instance then fuzzes the synthesized interface. If the bug is triggered, the pipeline terminates successfully; otherwise, the run ends on timeout, coverage feedback is collected, and the agent uses that feedback to revise the harness for the next iteration.

\subsection{Implementation Details}
\label{sec:impl}

We implement our agents using the Model Context Protocol (MCP)~\cite{mcp}. The pipeline uses two agents. The \emph{generation agent} (i) synthesizes a fuzzing harness from the patch context, (ii) converts the harness into a Syzkaller pseudo-syscall, and (iii) generates the corresponding \texttt{syzlang} syscall description. The \emph{repair agent} is lightweight and is invoked only when integration fails to build; it consumes Syzkaller compilation logs and fixes errors introduced by newly added pseudo-syscall code or syscall descriptions.

The conversion into pseudo-syscall code follows a set of pseudo-syscall conventions distilled from existing Syzkaller pseudo-syscalls. We implement purely syntactic and deterministic edits (e.g., required wrapper macros and boilerplate) via scripts. For transformations that require semantic awareness of argument usage—most notably, inserting explicit type casts—we rely on the generation agent. This is necessary because pseudo-syscall arguments are passed using Syzkaller’s executor conventions (e.g., as \texttt{volatile long} values) and must be cast to the intended concrete types at the start of the pseudo-syscall body before they are used.

We implement an MCP server that exposes the tools required by the agents (listed below), including kernel code navigation (e.g., symbol lookup, pattern search and targeted source retrieval) and compilation checks. The agents are driven via Codex CLI with a fixed system prompt (\texttt{AGENTS.md}) and task-specific user prompts (e.g., harness synthesis, pseudo-syscall conversion, and syscall-description generation). The repair agent is constrained to edit only the newly injected code blocks and is capped at a small number of repair rounds (five in our implementation). The concrete task prompts are listed in Appendix~\ref{app:prompts}.

\vspace{-.15in}
\section{Evaluation}
\label{sec:eval}

\subsection{Experiment Setup}
\label{sec:setup}
For each target patch, the pipeline finds the kernel revision immediately preceding the patch commit and builds that kernel using GCC-11. We use the Syzbot kernel configuration~\cite{syzbot}, which provides broad subsystem coverage and enables KASAN to detect memory safety violations at the time they are triggered. We use GPT-5.2-codex with xhigh reasoning as the primary LLM.

We ran the fuzzing experiments on a dedicated server equipped with two AMD EPYC 7543 32-core processors
(128 hardware threads in total) and 944 GiB of RAM. The fuzzing workload was distributed across 4 virtual machines, each provisioned with 8 vCPUs, for a total of 32 concurrent fuzzing threads.

A kernel crash alone is insufficient to establish reproduction of the
vulnerability fixed by the target patch: a generated harness may expose
an unrelated bug along the same execution path. We therefore use a
patch-differential oracle for all candidate reproductions.

For each generated PoC, we execute the same input under the same kernel
configuration and repetition budget on both the vulnerable pre-patch
kernel and the corresponding patched kernel. We count a target as
patch-validated only if (i) the PoC reproducibly triggers a kernel
failure on the pre-patch kernel, (ii) the corresponding failure is no
longer observed on the patched kernel, and (iii) the crash trace or
execution path is consistent with the function or root cause affected
by the patch. Ambiguous cases are manually inspected and conservatively
excluded.

\subsection{Experiment Metrics}
In our evaluations, we care about both \emph{effectiveness} and \emph{efficiency}. Effectiveness is measured by the \emph{success rate} (SR), i.e., the fraction of target vulnerabilities for which the system successfully triggers a kernel crash within the allowed budget. Efficiency is measured by the \emph{time-to-reproduction} (TTR), i.e., the end-to-end wall-clock time from the start of bug reproduction to the first crash. Over successful cases, we report both the \emph{mean time-to-reproduction} (MTTR) and the \emph{median time-to-reproduction} (MedTTR), since bug reproduction time can exhibit a long-tailed distribution. We also report the \emph{first-iteration success rate} (FISR), which captures how often the initial harness is sufficient without iterative refinement.

\subsection{Evaluation Datasets}
\label{subsec:dataset}

We evaluate \system on multiple complementary datasets.

\paragraph{KernelCTF patch dataset (overall performance)}
Our primary dataset is derived from Google’s KernelCTF~\cite{kernelctf}, a vulnerability rewards program that includes real-world Linux kernel 0-day and 1-day vulnerabilities with demonstrated exploitation. These cases provide strong ground truth for bug reproduction because they are \emph{known to be triggerable}, unlike general CVE corpora that may include reports that are difficult or impossible to reproduce in practice~\cite{linuxkernelcves}. Many KernelCTF cases also provide public documents (after disclosure grace periods) describing root causes, trigger conditions, and exploit strategies, which facilitate post-mortem analysis of failures. We include all cases released before the end of 2025 and retain only those whose fixing commits are available in the upstream Linux repository and whose vulnerable kernels boot successfully in our VM environment. This yields 100 cases.

\paragraph{SyzDirect dataset (comparative study)}
For head-to-head comparison against state-of-the-art directed greybox fuzzing for the Linux kernel (e.g., SyzDirect), we additionally evaluate \system on the dataset used by SyzDirect~\cite{tan2023syzdirect}. These cases originate from Syzbot~\cite{syzbot} and are similarly intended to be triggerable. Evaluating on the same benchmark controls dataset effects and enables a fair comparison under identical targets.
This dataset contains 100 cases.

\paragraph{Ablation subset}
For ablation studies, we randomly sample 30\% of the cases from each dataset, forming 60 unique bugs in total.

\paragraph{Recent syzbot patch dataset (recent known-triggerable bugs)}
To evaluate \system on recent real-world bugs with clearly interpretable triggerability ground truth, we collect 50 syzbot UAF/OOB bugs fixed after March 2026. Each target has an existing syzbot crash or reproducer and a corresponding public fix, independently establishing that the bug is triggerable. During our evaluation, the public reproducer is withheld from \system and used only as ground truth; \system receives only the fix commit. This dataset complements KernelCTF with recent Linux kernel bugs while avoiding the unknown-triggerability limitation of the patch-stream dataset below.

\paragraph{Unlabeled patch dataset}
To evaluate \system beyond curated benchmarks, we further test it on \emph{unlabeled} patches without publicly available PoCs or CVEs. We collect candidate patches submitted in 2026 and use an LLM-assisted filtering step to retain patches likely to fix bugs with high security impact and reproducible in our environment. We apply the following rules to use LLMs to identify promising patches: (i) patches fixing UAF/OOB bugs, (ii) bugs whose vulnerable code is compiled (according to the syzbot config we use for testing) and feasible to reach (not dead code), (iii) bugs whose vulnerable code is reachable via syscalls to unprivileged users. Overall, we obtain 64 patches.

Unlike the datasets above, these patches have no independent triggerability ground truth. We therefore use this dataset to measure operational yield in a patch-only setting, rather than treating 64 as a denominator of known-triggerable bugs.

\begin{table}[t]
\centering
\small
\begin{tabular}{cccc}
\toprule
SR  & Mean TTR & Med.\ TTR \\
\midrule
78\% & 2.9\,h & 1.2\,h \\
\bottomrule
\end{tabular}
\caption{Overall performance of \system on the KernelCTF dataset. SR = success rate; TTR = time-to-reproduction (wall-clock, successful cases).}
\vspace{-.1in}
\label{table:kernelctf}
\end{table}

\begin{table}[t]
\centering
\small
\begin{tabular}{lcc}
\toprule
Component & Mean & Median \\
\midrule
Harness generation agent & 0.5\,h  & 0.3\,h  \\
Compilation repair agent & 2\,min  & 0       \\
Syzkaller fuzzing        & 2.2\,h  & 27\,min \\
\bottomrule
\end{tabular}
\caption{Time breakdown of \system over successful KernelCTF cases.}
\label{table:timekernelctf}
\vspace{-.3in}
\end{table}

\begin{table}[t]
\centering
\small
\begin{tabular}{lcccc}
\toprule
 & SR & 1st-iter SR & Mean TTR & Med TTR \\
\midrule
UAF   & 79.0\%  & 43.8\% & 3.1\,h  & 1.5\,h  \\
OOB & 82.4\%  & 71.4\%     & 1.9\,h & 0.5\,h \\
\bottomrule
\end{tabular}
\caption{Performance of \system on different bug types. SR = success rate; TTR = time-to-reproduction (wall-clock, successful cases).}
\label{table:type}
\vspace{-.3in}
\end{table}

\subsection{Evaluation results}
We organize our evaluation around three questions:
\squishlist
\item \textbf{RQ1}: How effective and efficient is \system on real-world triggerable Linux kernel vulnerabilities?
\item \textbf{RQ2}: How does \system compare with prior state-of-the-art directed greybox fuzzing?
\item \textbf{RQ3}: What do ablation studies reveal about the
division of labor between LLM-recovered trigger scaffolding and fuzzer-driven
concrete-value search, and how stable is \system across repeated runs
 and unlabeled datasets?
\squishend

\subsubsection{Overall performance}

\begin{figure}
\centering
  \includegraphics[width=0.35\textwidth]{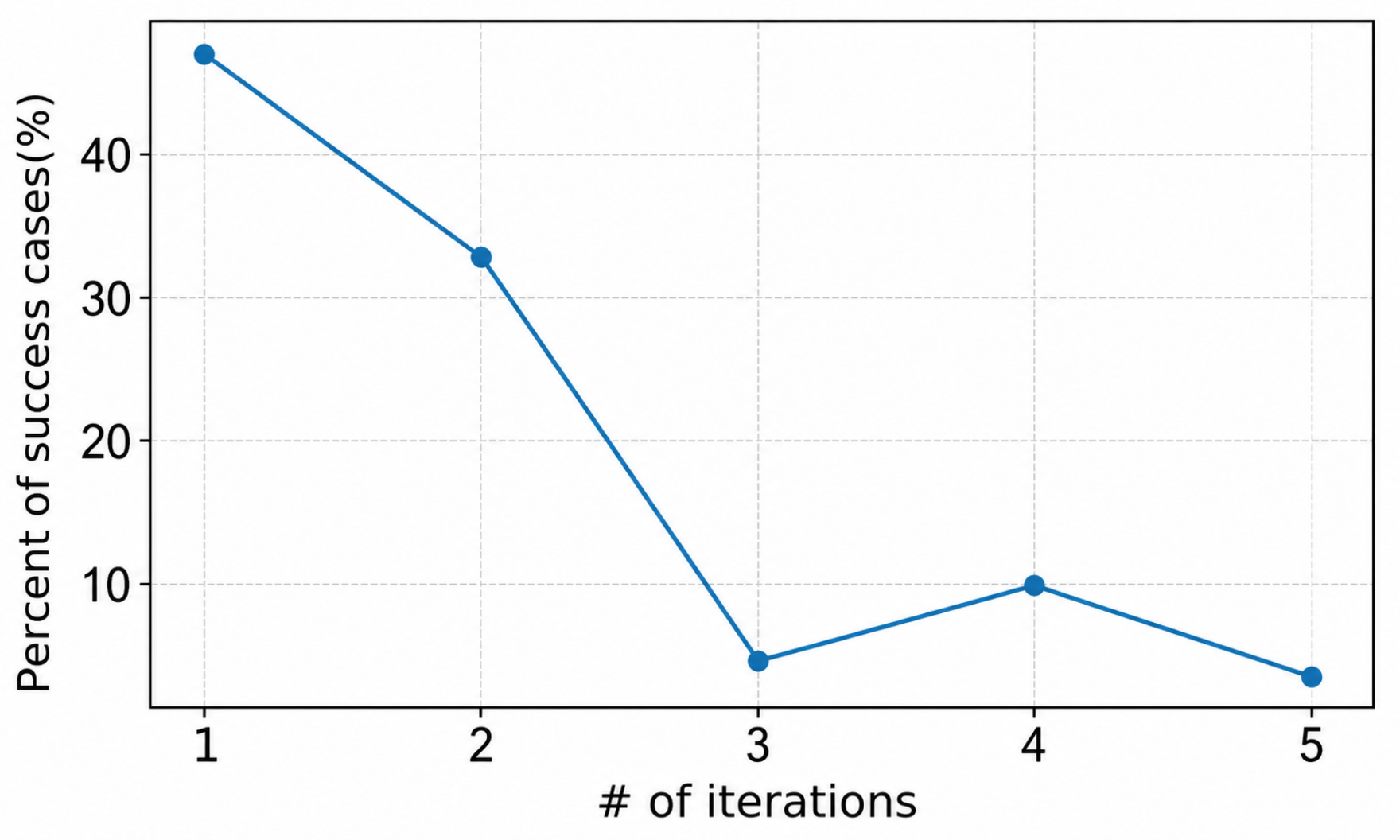}
  \caption{The percentage of successful cases in each iteration.}
  \label{fig:iterdistribution}
\vspace{-.2in}
\end{figure}

\autoref{table:kernelctf} summarizes the effectiveness and efficiency of \system on the KernelCTF dataset. \system initially generates candidate reproductions for 80/100 targets. After applying the patch-differential validation described in \S\ref{sec:setup}, 78/100 (78\%) are confirmed as reproductions of the vulnerabilities fixed by the corresponding patches. Among the 78 validated successful cases, 48.72\% are triggered in the first iteration (out of at most five iterations), while iterative refinement substantially improves success for the remaining long-tail cases. For successful cases, the mean time-to-reproduction is 2.9 h and the median is 1.2 h, suggesting that many cases reproduce quickly while a smaller subset requires longer fuzzing and/or further refinement.

\PP{Time breakdown} To better understand where time is spent, we break down the end-to-end cost into three components: (i) the \emph{harness generation agent}, which crafts the pseudo-syscall for triggering the bug and generates the corresponding syscall description; (ii) the \emph{compilation repair agent}, which is invoked only when integration introduces Syzkaller build errors; and (iii) Syzkaller fuzzing time, during which Syzkaller mutates the exposed harness parameters.

\autoref{table:timekernelctf} shows that the agent's overhead is modest relative to fuzzing. The harness generation agent requires 0.5 hours on average (0.3-hour median), even accounting for multi-iteration refinement. The compilation repair agent is rarely needed: its median time is zero, indicating that at least half of successful cases integrate without any build errors. Syzkaller accounts for most of the wall-clock time (2.2-hour mean), reflecting the inherently stochastic search over concrete values and runtime effects. Notably, this fuzzing time does not entail LLM API cost, but the agent time includes that.

\PP{Success rate by iteration}
\autoref{fig:iterdistribution} breaks down \emph{successful} cases by iteration. Successful bug reproductions are strongly front-loaded: 48.72\% of the successes occur in the first iteration and 34.62\% in the second, meaning that 83.34\% of all successful cases are reproduced within two iterations. The remaining 16.67\% form a clear long tail spread across later iterations (5.13\% in iteration 3, 7.69\% in iteration 4, and 3.85\% in iteration 5).

This distribution highlights two points. First, the high fraction of first-iteration successes shows that the initial harness is often  sufficient when it captures the required setup logic and parameterization—i.e., it fixes the prerequisite sequence correctly, assigns some key values properly, and exposes the remaining uncertain inputs for fuzzing. Second, the later-iteration successes show that iterative refinement remains useful for a few harder cases that are not reproduced in the short period, even though they eventually become reproducible under revised harnesses and additional fuzzing budget.

\PP{Runtime and training-data leakage analysis.}
Because \system leverages an LLM agent, a natural concern is that reproduction success may be inflated by access to public PoCs or vulnerability information, either through \emph{runtime retrieval} or \emph{training-data memorization}. We evaluate these two threats separately.

First, to examine runtime retrieval, we manually inspect the complete agent logs from 30 randomly selected KernelCTF cases. We check whether the agent invokes web search, accesses KernelCTF writeups, or retrieves public reproducer code. We find no evidence of any such behavior in these 30 cases. This analysis provides evidence that the observed reproduction results are not driven by the agent directly retrieving public solutions during execution.

Second, we investigate potential training-data leakage through a \emph{knowledge-cutoff analysis}. We partition KernelCTF cases according to whether the corresponding patch was submitted before or after the model's stated knowledge cutoff and compare reproduction success rates between the two groups.

The cutoff date of GPT-5.2-codex (August~31,~2025~\cite{gpt52codex}) yields only 8 KernelCTF cases after the cutoff, which is too small for a balanced comparison. Hence, we additionally evaluate \system with GPT-5.1-codex-max (cutoff: September~30,~2024~\cite{gpt52codex}), for which 58\% of KernelCTF cases fall before the cutoff and 42\% after.
Under GPT-5.1-codex-max, \system achieves a 70\% patch-validated reproduction success rate; the success rate is 70.69\% on pre-cutoff cases and 69.05\% on post-cutoff cases. The negligible difference suggests that \system’s effectiveness is unlikely to be driven by memorization of post-cutoff patches or writeups, but instead by its ability to infer prerequisite structure from patch-local context and reliance on fuzzing for concrete-value discovery.

\PP{Performance by bug type}
We seek to understand the effect of bug types on performance. 98\% of the KernelCTF cases are use-after-free or memory out-of-bounds access bugs, since they constitute the majority of exploitable types. UAF bugs typically depend on object lifetime management (freeing, reallocation, and subsequent stale use), whereas OOB bugs are often triggered by violating explicit size/bounds constraints. Different patterns may have different complexities, causing differences in performance. We label each KernelCTF case as UAF or OOB and report the breakdown in \autoref{table:type}. Overall, \system achieves comparable patch validation success rates across the two classes, with UAF being slightly lower than OOB (79.0\% vs. 82.4\%). However, the efficiency gap is pronounced. For OOB cases, 71.4\% of the successful bug reproductions occur in the first iteration, compared to 43.8\% for UAF. Moreover, successful UAF cases require substantially more time (mean TTR 3.1~h vs.\ 1.9~h; median TTR 1.5~h vs.\ 0.5~h).

\PP{Performance on race-condition bugs}
Race- and timing-sensitive bugs pose a different challenge from other bug types: even with the correct trigger scaffold and concrete values, reproduction may require a particular execution interleaving. To directly evaluate such cases, we manually inspect all 100 KernelCTF targets and label whether triggering requires race/timing-sensitive interleavings. We identify 24 race-condition cases and 76 non-race cases. After validation, \system reproduces 19/24 race cases (79.2\%) and 59/76 non-race cases (77.6\%). These results demonstrate that, race-condition bugs do not exhibit a lower reproduction success rate in our evaluation. Nevertheless, timing sensitivity can increase reproduction latency because the required interleaving may only occur after repeated executions.

\PP{Token cost analysis}
Because \system relies on an LLM agent, an important practical question is whether its API cost is prohibitive. On GPT-5.2-codex, the average token cost of \system is approximately \$2 per case, compared with \$2.50 for the LLM-only baseline, corresponding to about a 20\% reduction in LLM inference cost. This cost is modest relative to the overall bug reproduction pipeline, especially because the dominant component of wall-clock time is Syzkaller fuzzing rather than LLM interaction. In other words, \system primarily uses the LLM to recover trigger scaffolding and synthesize/refine the harness, while the expensive exploration over concrete values is offloaded to fuzzing at essentially no additional API cost. This makes the overall design practical for large-scale patch-based bug reproduction.

\PP{Failure analysis}
To understand the remaining gaps, we manually inspected unsuccessful cases and grouped failures into three recurring categories. The dominant failure mode is \emph{incorrect scaffolding} (14 cases out of 20): fuzzing with the synthesized harness reaches the correct patched file or function but fails to satisfy the vulnerability's triggering conditions, either because it omits a required operation sequence or because it constructs an incompatible combination of object types and resources. This suggests that, for a subset of vulnerabilities, LLMs still fail to provide the appropriate trigger conditions, even when we allowed multiple iterations (e.g., after fuzzing feedback).

The remaining failures are driven by \emph{dynamic} conditions that are harder to encode in a static harness. In 4 out of 20 cases, the harness reaches the relevant code path but fails to hit a required race or timing window. This could be either (1) the scaffolding not allowing sufficient concurrency, contention, or scheduling variability, and (2) it did, but fuzzing failed to find the right concrete values to hit the vulnerable scheduling window, e.g., wrong values for \texttt{usleep()}. Given the complexity of such timing-related issues, we defer further investigation into resolving such cases for future work.

Finally, 2 out of 20 cases exhibit \emph{minor issues in the scaffolding}: the harness performs very similar operations to the ground truth but with ordering or parameter discrepancies that ultimately prevent the bug from triggering. These cases do contain the necessary syscalls and structures, but the parameter constraints, constants set by the harness, or the order of syscalls are incorrect.

\subsubsection{Comparative study}

We conduct a controlled head-to-head comparison between \system and
SyzDirect~\cite{tan2023syzdirect} on the 100-case SyzDirect benchmark.
We rerun the public SyzDirect implementation using the same vulnerable
kernel revisions and configurations, Syzkaller revision, compiler,
hardware allocation, 24-hour per-target budget, trial setting as \system. We also
apply the same patch-differential reproduction oracle to both systems,
such that a target is counted as successfully reproduced only when the
generated PoC triggers the vulnerable pre-patch kernel, no longer triggers
the corresponding failure on the patched kernel, and is consistent with
the patched vulnerability.

Under these matched settings, \system achieves a 73\% patch-validated
reproduction rate, compared with 25\% for SyzDirect.

The 25\% result differs from the 42\% reported in the original SyzDirect
paper because the two evaluations use different trial aggregation.
SyzDirect's reported 42\% counts a target as reproduced if at least one
of its ten independent trials succeeds, whereas our controlled comparison
evaluates both systems under the same single-run trial setting and resource budget.
We therefore use the 25\% result from our controlled rerun, rather than
SyzDirect's published 42\%, as the baseline for comparison with \system.

Thus, under matched experimental conditions and a common patch-sensitive
reproduction oracle, \system reproduces 73\% of the targets compared with
25\% for SyzDirect. These results support our design hypothesis that
explicitly recovering and fixing the patch-specific trigger scaffold
substantially reduces the search difficulty compared with SyzDirect's
directed-fuzzing formulation.

\subsubsection{Ablation Study, Stability, and Case Studies}
\label{subsec:ablation}

\system combines two key design choices that we ablate here: (i) delegating the search over uncertain, bug-critical concrete input values for syscalls to fuzzing rather than direct LLM generation, and (ii) iteratively refining the harness using hierarchical reachability feedback. We quantify the contribution of each component through an ablation study on a random 30\% subset of cases from the KernelCTF and SyzDirect datasets (\autoref{subsec:dataset}).

\PP{LLM-only generation (no fuzzing harness).}
We isolate the benefits of combining the LLM-derived fuzzing harness {\em and} the fuzzing-driven concrete search. We replace the harness synthesis with direct PoC generation: the agent emits a complete reproducer program (with all concrete values fixed) rather than a harness that exposes uncertainty to mutation.
Moreover, unlike fuzzing with an explicit time budget, an LLM-only approach has no natural stopping rule for this concrete search problem; the model itself effectively decides when to stop refining the program, even if the resulting PoC still fails to trigger the bug.
This LLM-only solution achieves 65\% SR, compared to 78.3\% with \system on the same ablation set. The gap is expected: while the agent can often infer trigger scaffolding (e.g., syscall ordering and protocol steps), it must also guess \emph{precise concrete values} that may lie in sparse or extreme regions and may interact with nondeterministic runtime effects.
More generally, closing this gap via repeated prompting would require many costly LLM trials over a large concrete space, whereas fuzzing can explore these uncertain parameters cheaply at high throughput. This ablation supports our design choice to use the LLM to identify appropriate trigger scaffolding while delegating concrete-value discovery and nondeterminism exploration to fuzzing.

\begin{table*}[]
\centering
\scriptsize
\setlength{\tabcolsep}{3pt}
\begin{tabular}{lccccccccccc}
\toprule
Variant & SR & Both w/ \system
& \multicolumn{2}{c}{Variant TTR on Both}
& \multicolumn{2}{c}{\system TTR on Both}
& \multicolumn{5}{c}{Targets Reproduced by Budget} \\
\cmidrule(lr){4-5}
\cmidrule(lr){6-7}
\cmidrule(lr){8-12}
& & & Mean & Med & Mean & Med
& $\le$0.5h & $\le$1h & $\le$2h & $\le$4h & $\le$8h \\
\midrule
1 iter (long fuzz)
& 44/60 (73.3\%) & 39
& 1.9\,h & 0.2\,h
& 1.1\,h & 0.4\,h
& 30 & 32 & 35 & 38 & 39 \\

3 iters w/ feedback
& 44/60 (73.3\%) & 42
& 1.3\,h & 0.5\,h
& 1.0\,h & 0.4\,h
& 25 & 31 & 38 & 40 & 42 \\

5 iters w/o feedback
& 46/60 (76.7\%) & 43
& 1.3\,h & 0.4\,h
& 1.0\,h & 0.4\,h
& 24 & 30 & 38 & 40 & 42 \\
\midrule

\system (5 iters w/ feedback)
& 47/60 (78.3\%) & 47
& 1.1\,h & 0.5\,h
& 1.1\,h & 0.5\,h
& 27 & 34 & 41 & 45 & 46 \\
\bottomrule
\end{tabular}
\caption{Ablation of iteration mechanisms on the ablation subset. TTR is measured excluding kernel build time. \textbf{Both w/ \system} is the number of targets reproduced by both the variant and
\system. \textbf{Variant TTR on Both} and \textbf{\system TTR on Both} report the variant's and \system's TTR, respectively, on the targets reproduced by both, enabling a paired comparison. Budget columns count reproduced targets among all 60 cases.}
\label{table:ablation-iter}
\vspace{-.35in}
\end{table*}

\PP{Iteration and feedback mechanisms.}
Finally, we study the impact of iterative refinement. \system performs up to five iterations; after a failed iteration, it extracts hierarchical reachability feedback (file/function/line) and
uses it to revise the next harness. \autoref{table:ablation-iter} compares four variants under a fixed end-to-end budget of approximately 24 hours. To avoid conflating reproduction speed with
differences in the set of successfully reproduced targets, we report TTR only on targets reproduced by both the variant and \system. We also report the number of targets reproduced within fixed time
budgets.

The results indicate that hierarchical feedback mainly benefits \emph{efficiency}, while providing little evidence of improved final effectiveness. The cleanest comparison is between five
iterations without feedback and \system. Their final validated success counts are nearly identical: 46/60 versus 47/60. However, on the 43 targets reproduced by both variants, \system reduces mean TTR
from 1.3h to 1.0h, while median TTR remains comparable at 0.4h. More importantly, \system reproduces more targets at intermediate budgets: 34 versus 30 within 1h, 41 versus 38 within 2h, and
45 versus 40 within 4h. Thus, feedback does not substantially increase the final number of reproduced targets in this single-run ablation, but it helps reproduce many targets earlier.

Iterations alone also help, but less consistently. A single long fuzzing run reproduces 44/60 targets, while three feedback-guided iterations also reproduce 44/60 targets but shift more successes
into earlier budgets after the initial short-fuzz phase. Five iterations without feedback improves final SR to 46/60, showing that additional attempts can recover some failures even without
structured guidance. From the 1h budget onward, the full \system configuration achieves the strongest fixed-budget profile. Overall, these results suggest that hierarchical feedback is best
viewed as an efficiency mechanism: it helps identify inadequate harness scaffolds and steer revisions earlier, rather than providing strong evidence of a large improvement in final
reproduction effectiveness.

\PP{Stability across repeated runs}
A practical question for \system is how sensitive it is to pipeline
stochasticity. Variability can arise from two sources: (i) LLM-driven
harness synthesis is nondeterministic and may produce different scaffolds
across executions, and (ii) Syzkaller fuzzing is probabilistic due to
randomized mutation and scheduling.

To characterize this variability more systematically, we independently
rerun the full end-to-end pipeline five times on the 60-case ablation
subset under the same configuration and per-target budget. Each run starts
from the same target patch and configuration, but performs fresh harness
synthesis and fuzzing. We report both aggregate reproduction rates across
runs and per-target success frequencies.

Across the five independent runs, \system reproduces 47/60, 52/60,
47/60, 50/60, and 50/60 targets, respectively. The success rate ranges
from 78.3\% to 86.7\%, with a mean of 49.2/60 (82.0\%). Thus, aggregate
reproduction success remains within a relatively narrow range across
independent executions.
\begin{figure}
\centering
  \includegraphics[width=0.4\textwidth]{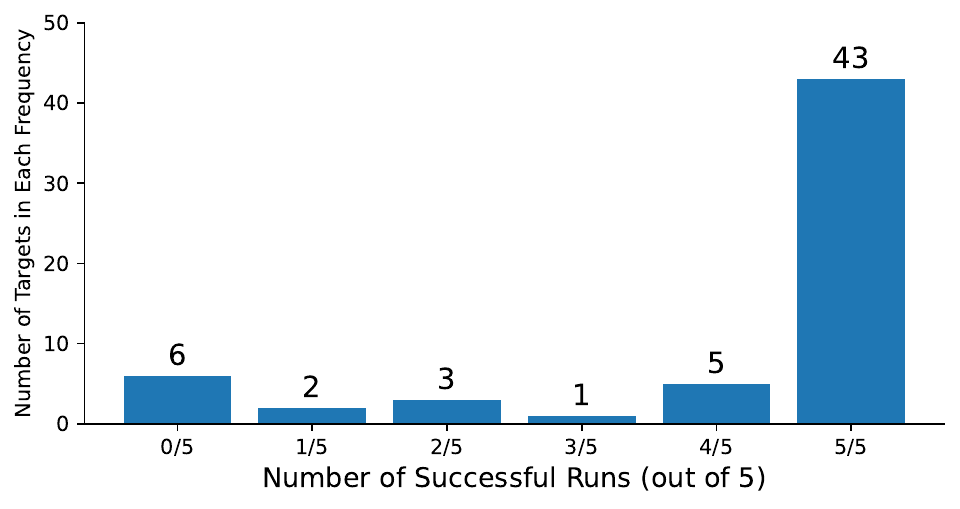}
  \caption{Distribution of per-target reproduction frequency across five
independent runs on the 60-case ablation subset.
Most targets are consistently reproducible across runs, while
run-to-run variability is concentrated in a small subset of cases.}
  \label{fig:stability}
\vspace{-.35in}
\end{figure}
The per-target success distribution in \autoref{fig:stability} provides
a more detailed view. Among the 60 targets, 43 are reproduced in all
five runs, 5 in four runs, 1 in three runs, 3 in two runs, 2 in one run,
and 6 in none of the runs. Thus, 48/60 targets are reproduced in at
least four of five runs, while 54/60 are reproduced at least once.
Most run-to-run variation is therefore concentrated in a relatively
small set of marginal targets.

Overall, the five-run analysis shows that \system has relatively stable
aggregate reproduction rates, while stochasticity is concentrated in a
limited subset of targets rather than affecting all cases uniformly.
We therefore use this repeated-run distribution to characterize
stochastic uncertainty and avoid interpreting small differences between
system variants as meaningful when they are comparable to the observed
run-to-run variation.

\begin{figure}[t]
    \centering
    \begin{minipage}[t]{0.98\linewidth}
\begin{minted}[
    fontsize=\scriptsize,
    breaklines,
    frame=single,
    framesep=1.5mm,
    bgcolor=codebg,
    tabsize=2,
    escapeinside=||
]{diff}
net: sched: cls_u32: Undo tcf_bind_filter if u32_replace_hw_knode

When u32_replace_hw_knode fails, we need to undo the tcf_bind_filter
operation done at u32_set_parms.


diff --git a/net/sched/cls_u32.c b/net/sched/cls_u32.c
index d15d50de79802..ed358466d042a 100644
--- a/net/sched/cls_u32.c
+++ b/net/sched/cls_u32.c
@@ -1074,15 +1088,18 @@ static int u32_change(struct net *net, struct sk_buff *in_skb,
 	}
 #endif
 
-	err = u32_set_parms(net, tp, base, n, tb, tca[TCA_RATE],
+	err = u32_set_parms(net, tp, n, tb, tca[TCA_RATE],
 			    flags, n->flags, extack);
+
+	u32_bind_filter(tp, n, base, tb);
+
 	if (err == 0) {
 		struct tc_u_knode __rcu **ins;
 		struct tc_u_knode *pins;
 
 		err = u32_replace_hw_knode(tp, n, flags, extack);
 		if (err)
-			goto errhw;
+			goto errunbind;
 
 		if (!tc_in_hw(n->flags))
 			n->flags |= TCA_CLS_FLAGS_NOT_IN_HW;
@@ -1100,7 +1117,9 @@ static int u32_change(struct net *net, struct sk_buff *in_skb,
 		return 0;
 	}
 
-errhw:
+errunbind:
+	u32_unbind_filter(tp, n, tb);
+
 #ifdef CONFIG_CLS_U32_MARK
 	free_percpu(n->pcpu_success);
 #endif
\end{minted}
    \end{minipage}
    \caption{A case that \system succeeds, while the LLM-only method fails; the patch is simplified due to space}
    \label{fig:case2}
    \vspace{-.3in}
\end{figure}

\PP{Case studies}
We next examine two representative cases to understand when \system succeeds, especially where an LLM-only approach fails. The first revisits the motivating SCSI example from \autoref{fig:motivateexample}, now from the perspective of empirical reproduction.
As discussed earlier, the patch identifies \texttt{cmd\_len} as the critical field but does not reveal the exact triggering value. The LLM-only approach captures this high-level insight yet fixes \texttt{cmd\_len} to the non-triggering value \texttt{5}. By contrast, \system leaves \texttt{cmd\_len} exposed for fuzzing and discovers the actual trigger value \texttt{0}. This case illustrates a recurring pattern: the LLM can often localize the right field, while fuzzing is needed to recover the precise concrete value.

The second example, shown in \autoref{fig:case2}, is a use-after-free bug in the \texttt{cls\_u32} traffic-control subsystem. The patch shows that when \texttt{u32\_replace\_hw\_knode()} fails, the kernel must explicitly undo the prior \texttt{u32\_bind\_filter()} operation; otherwise, later cleanup can dereference stale class/filter state. Reproducing this bug therefore requires more than reaching the correct subsystem: the reproducer must construct a precise traffic-control configuration, bind the filter to the intended class, drive execution down the hardware-replace failure path, and then trigger the subsequent cleanup sequence. In our pipeline, the LLM successfully synthesizes this high-level scaffold, while the harness exposes 12 bug-critical arguments for fuzzing, including class IDs, \texttt{u32} flags, hash-table ID, node ID, priority, protocol, quantum, and the choice of whether to take the replace path. These parameters jointly determine whether the kernel enters the relevant error path and whether later cleanup dereferences stale state. The LLM-only approach does not reliably identify a working combination of these values and fails to generate a working PoC, whereas \system succeeds by combining LLM-derived structural setup with high-throughput search over the remaining concrete space.

\subsubsection{Recent known-triggerable syzbot bugs}

To evaluate \system on recent real-world vulnerabilities with a clearly interpretable denominator, we conduct an additional experiment on 50 syzbot UAF/OOB bugs fixed after March 2026. Each target has an existing syzbot crash or reproducer and a corresponding public fix, independently establishing that the bug is triggerable. During evaluation, we withhold the public reproducer and provide \system only with the fix commit.

\system reproduces 40 of the 50 targets (80\%) after validation. Because every target in this dataset is independently known to be triggerable, this result provides a clearly interpretable success rate on recent real-world Linux kernel bugs and complements the older KernelCTF benchmark with substantially more recent fixes.

\subsubsection{Unlabeled patches}

Unlike the preceding datasets, the bugs from the unlabeled patch dataset (described in \S\ref{subsec:dataset}) have no independent triggerability ground truth. We therefore do not interpret this experiment as a reproduction success rate over known-triggerable bugs. Instead, it measures the operational yield of applying \system to a recent patch stream in which only the patch is initially available.

Of the 64 unlabeled patches, \system successfully reproduces 29 and fails on 35,
yielding a validated success rate of 45.3\%. This is lower than the 80\% success rate on
the KernelCTF dataset. We attribute this gap to two likely factors. First,
unlike KernelCTF cases, the unlabeled patches are not externally validated as
triggerable vulnerabilities; some may correspond to bugs that are difficult to
trigger in our environment or are not triggerable at all. Second, these patches
span a broader set of kernel subsystems, including less commonly exercised
modules whose setup logic, object dependencies, and triggering conditions may
be less visible from public examples and documentation.
These factors make it harder for \system to recover the trigger scaffolding needed to reach the vulnerable path. To better understand these failures, we randomly sample 10 of the 35 unsuccessful cases for manual analysis. We find that 8 of the 10 are in fact not triggerable in our setup because necessary kernel configurations are missing; examples include \texttt{CONFIG\_MT7925E} for MediaTek MT7925E PCIe support and \texttt{CONFIG\_DRM\_AMDGPU} for AMD GPU support. As a comparison, KernelCTF cases are concentrated in a narrower set of commonly exercised kernel subsystems such as \texttt{net} and \texttt{netfilter}.

We further sampled 10 of these bugs to understand their true exploitability. According to KASAN bug reports, 6 are UAF read, 2 are UAF write, 1 is heap OOB write, and 1 is stack OOB read.
The three bugs with write primitives are highly likely to be exploitable.
However, given that we know the very first bug impact is not always the most severe~\cite{syzscope}, we further analyzed the subsequent impact of these bugs. It turns out that 4 of the UAF read actually have additional write primitives, including UAF write, OOB write, and arbitrary address write.
Take patch commit 424e95d62110 as an example, the first UAF read reported by KASAN is
\texttt{if (ae \&\& cf->data[0] != so->opt.rx\_ext\_address)} where \texttt{so} is the dangling pointer. However, subsequently there is also a UAF write using the same dangling pointer: \texttt{so->rx.buf[so->rx.idx++] = cf->data[i];}
Another example is patch commit 190a8c48ff62. The initial UAF read is on \texttt{mpol->nodes} (where \texttt{mpol} is the dangling pointer). The field is a bitmask that decides the offset of a subsequent array access \texttt{futex\_queues[node]}, leading to OOB read, retrieving a pointer of \texttt{struct futex\_hash\_bucket *}. The bogus pointer in OOB memory is then used to perform further dangerous memory-write operations, effectively leading to an arbitrary address write primitive.

\section{Discussion and Future Work}
\label{sec:discussion}

\subsection{Beyond fuzzing-based concrete search}
In this work, we use coverage-guided fuzzing as the backend for exploring
uncertain parameters and runtime effects. A natural extension is to combine the
synthesized harness with symbolic execution or other constraint-solving
techniques. Rather than replacing fuzzing, symbolic or concolic execution could
serve as a complementary backend for cases whose triggers are dominated by
precise value constraints, by solving path conditions over bug-critical harness
parameters more directly~\cite{stephens2016driller,yun2018qsym}. This is
particularly relevant because prior work has shown that hybrid fuzzing can
combine the fast exploration of fuzzing with the constraint-solving ability of
symbolic execution, including in the Linux kernel setting~\cite{kim2020hfl}.
At the same time, applying symbolic reasoning to kernel bug reproduction remains
challenging because of path explosion, complex environment modeling, implicit
syscall dependencies, and concurrency-sensitive runtime behavior~\cite{chipounov2011s2e,kim2020hfl}.
Exploring hybrid backends that combine LLM-generated harnesses with fuzzing and
symbolic reasoning is therefore a promising direction for future work.

\subsection{Generality beyond the Linux kernel}

Our evaluation focuses on Linux kernel patch-based vulnerability reproduction, consistent with prior work such as SyzDirect~\cite{tan2023syzdirect}. This focus is well motivated: the Linux kernel is one of the largest and most impactful open-source systems, underpinning cloud platforms, servers, embedded devices, and mobile systems. Its scale, complexity, and security relevance make it a compelling and challenging target for automated vulnerability reproduction.

Although our implementation targets the reproduction of Linux kernel bugs, the underlying
decomposition is generalizable to other targets. We envision it to apply to complex and stateful targets. For example, triggering bugs in
network protocols, REST APIs, and file systems often follows a similar
pattern: first, establish a valid internal state through a sequence of
interactions, then search for the specific values or operations that expose
the fault~\cite{pham2020aflnet,atlidakis2019restler,xu2019janus}. Although generic, applying our insight to other stateful targets would require
target-specific harness interfaces, which we leave for future work.

\section{Related Work}

\subsubsection{Directed greybox and kernel fuzzing}
Since AFLGo~\cite{bohme2017directed}, directed greybox fuzzing (DGF) has
evolved through more precise static guidance and target-distance estimation
(e.g., Hawkeye~\cite{chen2018hawkeye}, WindRanger~\cite{du2022windranger},
SelectFuzz~\cite{luo2023selectfuzz}). More recent work augments purely
structural guidance with richer pruning and semantic signals, including
predicate-based progress characterization and LLM-informed relevance
estimation~\cite{zhu2025locus, wang2026attentiondistance}. For kernel targets,
the central challenges are stateful subsystems, syscall dependencies, and
incomplete interface knowledge. SyzDirect~\cite{tan2023syzdirect} studies
dependency-guided direction for kernel interfaces, while related systems address
complementary bottlenecks in
automatic interface modeling, learned syscall-sequence mutation, and
context-adaptive fuzzing~\cite{bulekov2023fuzzng, xu2024mock, lee2026rtcon}.
Some work also focuses on the post-fuzzing pipeline, such as crash impact
analysis, exploit construction, and automated
repair~\cite{zou2022syzscope, 10.1145/3719027.3744841, deng2025chainfuzz,
mathai2025crashfixer}. Our setting differs from these efforts in that \system focuses on
\emph{patch-specific bug reproduction}: it uses the LLM to synthesize a
\emph{parameterized harness} that fixes the prerequisite structure while
leaving only the remaining concrete search to the fuzzer.

\subsubsection{LLM-assisted fuzzing and vulnerability triggering}
LLMs have recently been used to help fuzzers overcome semantic barriers that
random mutation rarely crosses. Some work treats the model as a language-aware
input generator or mutator, synthesizing structured programs, protocol
messages, or format-preserving mutations for targets whose inputs must satisfy
rich syntactic and semantic constraints~\cite{deng_large_2023,
deng_large_2024, xia_fuzz4all_2024, meng_large_2024, yang_hybrid_2025,
zhang_llamafuzz_2025}. In these systems, the LLM improves the validity and
diversity of fuzzing inputs, but the generated artifact is still a seed, test
case, constraint solution, or mutator output rather than a persistent fuzzing
entry point.

Other systems use LLM reasoning to steer fuzzing toward target behaviors. They
combine LLMs with execution feedback, concolic analysis, firmware-oriented
analysis, crash triage, reachable-input generation, custom mutators, and
property-based tests~\cite{shiraishi2026pilot,
tu2026cottontail, ji2026firmagent, herter2025gptrace, xu_directed_2025,
zeng_pbfuzz_2025}. Within kernel fuzzing,
KernelGPT~\cite{yang2024kernelgptenhancedkernelfuzzing} and
SyzGPT~\cite{zhang2025unlocking} use LLMs to recover fuzzing knowledge such as
syscall dependencies and low-frequency syscall usage. These approaches improve
guidance or seed quality, but they still leave the fuzzer operating over its
ordinary input interface.

Closely related efforts study automated vulnerability reproduction. CVE-GENIE
reconstructs vulnerable environments and produces verifiable exploits from CVE
entries~\cite{ullah2025cvegenie}, while CyberGym benchmarks AI agents on
real-world vulnerability reproduction tasks~\cite{wang2025cybergym}. These
systems show that agentic LLMs can reason about vulnerability-triggering
conditions, but their output is typically a concrete exploit, PoC, or benchmark
attempt. \system instead targets \emph{patch-based Linux kernel bug reproduction}: the LLM synthesizes a parameterized harness that preserves the
patch-implied setup sequence, and Syzkaller explores only the remaining
uncertain concrete values and runtime nondeterminism.

\subsubsection{Fuzz harness and driver generation}
Harness generation addresses a more specific bottleneck: before a fuzzer can
exercise a library or API, someone must expose the relevant functionality
through an executable driver with valid setup, object construction, and call
ordering. Earlier systems reduce this manual effort by mining or reconstructing
API usage. FUDGE extracts fuzz drivers from existing client
code~\cite{babic2019fudge}, FuzzGen synthesizes library fuzzers from
whole-system analysis~\cite{ispoglou2020fuzzgen}, and Hopper interprets
API-usage programs to explore library call sequences without fixed
drivers~\cite{chen2023hopper}.

LLM-based systems attack the same bottleneck with code generation and iterative
repair. OSS-Fuzz-Gen generates and repairs fuzz targets within the OSS-Fuzz
infrastructure~\cite{ossfuzzgen, liu2023aipoweredfuzzing}; Zhang \etal
characterize the effectiveness and failure modes of LLM-based fuzz-driver
generation~\cite{zhang2024llmfuzzdriver}. Subsequent work improves robustness
by adding coverage-guided prompt mutation~\cite{lyu2024promptfuzz},
code-knowledge graphs~\cite{xu2024ckgfuzzer}, structured code/documentation/API
knowledge~\cite{liu2025promefuzz}, tool-augmented retrieval and compilation
repair~\cite{yang2025harnessagent}, or binary-analysis context for black-box
libraries~\cite{hardgrove_liblmfuzz_2025}.

Existing harness-generation systems mostly
target user-space libraries or API functions and aim to improve broad coverage
or discover new bugs through standalone LibFuzzer/OSS-Fuzz-style drivers.
\system, in contrast, targets vulnerability reproducation in the Linux kernel.
Its harness encodes syscall ordering and kernel state
construction, and exposes only uncertain bug-relevant parameters, and then is
translated into a Syzkaller pseudo-syscall plus a \texttt{syzlang} description.
The goal is thus not general driver synthesis, but patch-grounded
bug reproduction: use the harness to constrain kernel fuzzing to the state space
implied by the patch, while leaving concrete values to high-throughput mutation.

\vspace{-.1in}
\section{Conclusion}

We designed \system for patch-based Linux kernel vulnerability reproduction. \system combines LLM-derived trigger scaffold with coverage-guided fuzzing by synthesizing a parameterized fuzzing harness that fixes the setup logic while exposing only bug-critical arguments for mutation. The harness is realized as a Syzkaller pseudo-syscall with a matching \texttt{syzlang} description. Our evaluations on KernelCTF and the SyzDirect benchmark show that \system achieve  achieves 78\% and 73\% bug reproduction success rates, respectively.

\bibliography{reference}

@String { APR          = {April} }

@String { COMPUTER     = {IEEE Computer Magazine} }

@String { DEC          = {December} }

@String { FEB          = {February} }

@String { JUL          = {July} }

@String { JUN          = {June} }

@String { MAY          = {May} }

@String { OCT          = {October} }

@inproceedings{ruan2024kernjc,
  title={Kernjc: Automated vulnerable environment generation for linux kernel vulnerabilities},
  author={Ruan, Bonan and Liu, Jiahao and Zhang, Chuqi and Liang, Zhenkai},
  booktitle={Proceedings of the 27th International Symposium on Research in Attacks, Intrusions and Defenses},
  pages={384--402},
  year={2024}
}

@Misc{gpt52codex,
  Title                    = {{GPT-5.2-codex}},
  HowPublished             = {\url{https://developers.openai.com/api/docs/models/gpt-5.2-codex}},
}

@article{zhang2025unlocking,
  title={Unlocking low frequency syscalls in kernel fuzzing with dependency-based rag},
  author={Zhang, Zhiyu and Li, Longxing and Liang, Ruigang and Chen, Kai},
  journal={Proceedings of the ACM on Software Engineering},
  volume={2},
  number={ISSTA},
  pages={848--870},
  year={2025},
  publisher={ACM New York, NY, USA}
}

@inproceedings{hao2022demystifying,
  title={Demystifying the dependency challenge in kernel fuzzing},
  author={Hao, Yu and Zhang, Hang and Li, Guoren and Du, Xingyun and Qian, Zhiyun and Sani, Ardalan Amiri},
  booktitle={Proceedings of the 44th International Conference on Software Engineering},
  pages={659--671},
  year={2022}
}

@inproceedings{tan2023syzdirect,
  title={Syzdirect: Directed greybox fuzzing for linux kernel},
  author={Tan, Xin and Zhang, Yuan and Lu, Jiadong and Xiong, Xin and Liu, Zhuang and Yang, Min},
  booktitle={Proceedings of the 2023 ACM SIGSAC Conference on Computer and Communications Security},
  pages={1630--1644},
  year={2023}
}

@inproceedings{zou2022syzscope,
  title={$\{$SyzScope$\}$: Revealing $\{$High-Risk$\}$ security impacts of $\{$Fuzzer-Exposed$\}$ bugs in linux kernel},
  author={Zou, Xiaochen and Li, Guoren and Chen, Weiteng and Zhang, Hang and Qian, Zhiyun},
  booktitle={31st USENIX security symposium (USENIX security 22)},
  pages={3201--3217},
  year={2022}
}

@inproceedings{hao2023syzdescribe,
  title={Syzdescribe: Principled, automated, static generation of syscall descriptions for kernel drivers},
  author={Hao, Yu and Li, Guoren and Zou, Xiaochen and Chen, Weiteng and Zhu, Shitong and Qian, Zhiyun and Sani, Ardalan Amiri},
  booktitle={2023 IEEE Symposium on Security and Privacy (SP)},
  pages={3262--3278},
  year={2023},
  organization={IEEE}
}

@Misc{mcp,
  Title                    = {{Introducing the Model Context Protocol
}},
  HowPublished             = {\url{https://www.anthropic.com/news/model-context-protocol}},
}

@misc{yang2024kernelgptenhancedkernelfuzzing,
      title={KernelGPT: Enhanced Kernel Fuzzing via Large Language Models}, 
      author={Chenyuan Yang and Zijie Zhao and Lingming Zhang},
      year={2024},
      eprint={2401.00563},
      archivePrefix={arXiv},
      primaryClass={cs.CR},
      url={https://arxiv.org/abs/2401.00563}, 
}

@inproceedings{syzscope,
  title={{SyzScope: Revealing High-Risk Security Impacts of Fuzzer-Exposed Bugs in Linux kernel}},
  author={Xiaochen Zou and Guoren Li and Weiteng Chen and Hang Zhang and Zhiyun Qian},
  booktitle={USENIX Security Symposium},
  year={2022}
}

@Misc{syzbot,
  Title                    = {{Syzbot}},
  Author                   = {Google},
  HowPublished             = {\url{https://syzkaller.appspot.com/upstream/}},
}

@misc{kernelctf,
	title = {KernelCTF},
	howpublished = {\url{https://google.github.io/security-research/kernelctf/rules.html}},
}

@Misc{linuxkernelcves,
  Title                    = {{CVEs}},
  HowPublished             = {\url{https://docs.kernel.org/process/cve.html}},
}

@inproceedings{deng2025chainfuzz,
  title={$\{$ChainFuzz$\}$: Exploiting Upstream Vulnerabilities in $\{$Open-Source$\}$ Supply Chains},
  author={Deng, Peng and Zhang, Lei and Meng, Yuchuan and Yang, Zhemin and Zhang, Yuan},
  booktitle={34th USENIX Security Symposium (USENIX Security 25)},
  pages={6199--6218},
  year={2025}
}

@inproceedings{zhu2025locus,
  title={Locus: Agentic predicate synthesis for directed fuzzing},
  author={Zhu, Jie and Shen, Chihao and Li, Ziyang and Yu, Jiahao and Chen, Yizheng and Pei, Kexin},
  booktitle={Proceedings of the IEEE/ACM 48th International Conference on Software Engineering (ICSE'26)},
  year={2026},
  publisher={ACM}
}

@inproceedings{luo2023selectfuzz,
  title={Selectfuzz: Efficient directed fuzzing with selective path exploration},
  author={Luo, Changhua and Meng, Wei and Li, Penghui},
  booktitle={2023 IEEE Symposium on Security and Privacy (SP)},
  pages={2693--2707},
  year={2023},
  organization={IEEE}
}

@inproceedings{du2022windranger,
  title={Windranger: A directed greybox fuzzer driven by deviation basic blocks},
  author={Du, Zhengjie and Li, Yuekang and Liu, Yang and Mao, Bing},
  booktitle={Proceedings of the 44th International Conference on Software Engineering},
  pages={2440--2451},
  year={2022}
}

@inproceedings{chen2018hawkeye,
  title={Hawkeye: Towards a desired directed grey-box fuzzer},
  author={Chen, Hongxu and Xue, Yinxing and Li, Yuekang and Chen, Bihuan and Xie, Xiaofei and Wu, Xiuheng and Liu, Yang},
  booktitle={Proceedings of the 2018 ACM SIGSAC conference on computer and communications security},
  pages={2095--2108},
  year={2018}
}

@article{mathai2025crashfixer,
  title={CrashFixer: A crash resolution agent for the Linux kernel},
  author={Mathai, Alex and Huang, Chenxi and Ma, Suwei and Kim, Jihwan and Mitchell, Hailie and Nogikh, Aleksandr and Maniatis, Petros and Ivan{\v{c}}i{\'c}, Franjo and Yang, Junfeng and Ray, Baishakhi},
  journal={arXiv preprint arXiv:2504.20412},
  year={2025}
}

@inproceedings{bohme2017directed,
  title={Directed greybox fuzzing},
  author={B{\"o}hme, Marcel and Pham, Van-Thuan and Nguyen, Manh-Dung and Roychoudhury, Abhik},
  booktitle={Proceedings of the 2017 ACM SIGSAC conference on computer and communications security},
  pages={2329--2344},
  year={2017}
}

@inproceedings{deng_large_2023,
	address = {New York, NY, USA},
	series = {{ISSTA} 2023},
	title = {Large {Language} {Models} {Are} {Zero}-{Shot} {Fuzzers}: {Fuzzing} {Deep}-{Learning} {Libraries} via {Large} {Language} {Models}},
	isbn = {979-8-4007-0221-1},
	shorttitle = {Large {Language} {Models} {Are} {Zero}-{Shot} {Fuzzers}},
	url = {https://dl.acm.org/doi/10.1145/3597926.3598067},
	doi = {10.1145/3597926.3598067},
	urldate = {2026-02-22},
	booktitle = {Proceedings of the 32nd {ACM} {SIGSOFT} {International} {Symposium} on {Software} {Testing} and {Analysis}},
	publisher = {Association for Computing Machinery},
	author = {Deng, Yinlin and Xia, Chunqiu Steven and Peng, Haoran and Yang, Chenyuan and Zhang, Lingming},
	month = jul,
	year = {2023},
	pages = {423--435},
}

@inproceedings{deng_large_2024,
	address = {New York, NY, USA},
	series = {{ICSE} '24},
	title = {Large {Language} {Models} are {Edge}-{Case} {Generators}: {Crafting} {Unusual} {Programs} for {Fuzzing} {Deep} {Learning} {Libraries}},
	isbn = {979-8-4007-0217-4},
	shorttitle = {Large {Language} {Models} are {Edge}-{Case} {Generators}},
	url = {https://dl.acm.org/doi/10.1145/3597503.3623343},
	doi = {10.1145/3597503.3623343},
	urldate = {2026-02-22},
	booktitle = {Proceedings of the {IEEE}/{ACM} 46th {International} {Conference} on {Software} {Engineering}},
	publisher = {Association for Computing Machinery},
	author = {Deng, Yinlin and Xia, Chunqiu Steven and Yang, Chenyuan and Zhang, Shizhuo Dylan and Yang, Shujing and Zhang, Lingming},
	month = feb,
	year = {2024},
	pages = {1--13},
}

@inproceedings{xia_fuzz4all_2024,
	address = {New York, NY, USA},
	series = {{ICSE} '24},
	title = {{Fuzz4All}: {Universal} {Fuzzing} with {Large} {Language} {Models}},
	isbn = {979-8-4007-0217-4},
	shorttitle = {{Fuzz4All}},
	url = {https://dl.acm.org/doi/10.1145/3597503.3639121},
	doi = {10.1145/3597503.3639121},
	urldate = {2026-02-22},
	booktitle = {Proceedings of the {IEEE}/{ACM} 46th {International} {Conference} on {Software} {Engineering}},
	publisher = {Association for Computing Machinery},
	author = {Xia, Chunqiu Steven and Paltenghi, Matteo and Le Tian, Jia and Pradel, Michael and Zhang, Lingming},
	month = apr,
	year = {2024},
	pages = {1--13},
}

@inproceedings{yang_hybrid_2025,
	title = {Hybrid {Language} {Processor} {Fuzzing} via {LLM}-{Based} {Constraint} {Solving}},
	isbn = {978-1-939133-52-6},
	url = {https://www.usenix.org/conference/usenixsecurity25/presentation/yang-yupeng},
	language = {en},
	urldate = {2026-02-22},
	author = {Yang, Yupeng and Yao, Shenglong and Chen, Jizhou and Lee, Wenke},
	year = {2025},
	pages = {6299--6318},
  booktitle = {34th {USENIX} {Security} {Symposium} ({USENIX} {Security} 25)},
  address = {San Diego, CA, USA},
  }

@inproceedings{meng_large_2024,
	address = {San Diego, CA, USA},
	title = {Large {Language} {Model} guided {Protocol} {Fuzzing}},
	isbn = {978-1-891562-93-8},
	url = {https://www.ndss-symposium.org/wp-content/uploads/2024-556-paper.pdf},
	doi = {10.14722/ndss.2024.24556},
	language = {en},
	urldate = {2026-02-22},
	booktitle = {Proceedings 2024 {Network} and {Distributed} {System} {Security} {Symposium}},
	publisher = {Internet Society},
	author = {Meng, Ruijie and Mirchev, Martin and Böhme, Marcel and Roychoudhury, Abhik},
	year = {2024},
}

@misc{zhang_llamafuzz_2025,
	title = {{LLAMAFUZZ}: {Large} {Language} {Model} {Enhanced} {Greybox} {Fuzzing}},
	shorttitle = {{LLAMAFUZZ}},
	url = {http://arxiv.org/abs/2406.07714},
	doi = {10.48550/arXiv.2406.07714},
	urldate = {2026-02-22},
	publisher = {arXiv},
	author = {Zhang, Hongxiang and Rong, Yuyang and He, Yifeng and Chen, Hao},
	month = oct,
	year = {2025},
	note = {arXiv:2406.07714 [cs]},
}

@misc{xu_directed_2025,
	title = {Directed {Greybox} {Fuzzing} via {Large} {Language} {Model}},
	url = {http://arxiv.org/abs/2505.03425},
	doi = {10.48550/arXiv.2505.03425},
	urldate = {2026-02-22},
	publisher = {arXiv},
	author = {Xu, Hanxiang and Zhao, Yanjie and Wang, Haoyu},
	month = may,
	year = {2025},
	note = {arXiv:2505.03425 [cs]},
}

@misc{zeng_pbfuzz_2025,
	title = {{PBFuzz}: {Agentic} {Directed} {Fuzzing} for {PoV} {Generation}},
	shorttitle = {{PBFuzz}},
	url = {http://arxiv.org/abs/2512.04611},
	doi = {10.48550/arXiv.2512.04611},
	urldate = {2026-02-22},
	publisher = {arXiv},
	author = {Zeng, Haochen and Bao, Andrew and Cheng, Jiajun and Song, Chengyu},
	month = dec,
	year = {2025},
	note = {arXiv:2512.04611 [cs]},
}

@inproceedings{hardgrove_liblmfuzz_2025,
	title = {{LibLMFuzz}: {LLM}-{Augmented} {Fuzz} {Target} {Generation} for {Black}-box {Libraries}},
	shorttitle = {{LibLMFuzz}},
	url = {http://arxiv.org/abs/2507.15058},
	doi = {10.1109/CARS67163.2025.11337309},
	urldate = {2026-02-22},
	booktitle = {2025 {Cyber} {Awareness} and {Research} {Symposium} ({CARS})},
	author = {Hardgrove, Ian and Hastings, John D.},
	month = oct,
	year = {2025},
	note = {arXiv:2507.15058 [cs]},
	pages = {1--6},
}

@inproceedings{10.1145/3719027.3744841,
author = {Zhang, Hao and Liu, Jian and Lu, Jie and Chen, Shaomin and Han, Tianshuo and Zhang, Bolun and Gong, Xiaorui},
title = {Reviving Discarded Vulnerabilities: Exploiting Previously Unexploitable Linux Kernel Bugs Through Control Metadata Fields},
year = {2025},
isbn = {9798400715259},
publisher = {Association for Computing Machinery},
address = {New York, NY, USA},
url = {https://doi.org/10.1145/3719027.3744841},
doi = {10.1145/3719027.3744841},
booktitle = {Proceedings of the 2025 ACM SIGSAC Conference on Computer and Communications Security},
pages = {1499–1513},
numpages = {15},
location = {Taipei, Taiwan},
series = {CCS '25}
}

@inproceedings{bulekov2023fuzzng,
  title={No Grammar, No Problem: Towards Fuzzing the Linux Kernel without System-Call Descriptions},
  author={Bulekov, Alexander and Das, Bandan and Hajnoczi, Stefan and Egele, Manuel},
  booktitle={Proceedings of the Network and Distributed System Security Symposium (NDSS 2023)},
  year={2023},
  publisher={The Internet Society},
  doi={10.14722/ndss.2023.24688},
  url={https://www.ndss-symposium.org/ndss-paper/no-grammar-no-problem-towards-fuzzing-the-linux-kernel-without-system-call-descriptions/}
}

@inproceedings{xu2024mock,
  title={MOCK: Optimizing Kernel Fuzzing Mutation with Context-aware Dependency},
  author={Xu, Jiacheng and Zhang, Xuhong and Ji, Shouling and Tian, Yuan and Zhao, Binbin and Wang, Qinying and Cheng, Peng and Chen, Jiming},
  booktitle={Proceedings of the Network and Distributed System Security Symposium (NDSS 2024)},
  year={2024},
  publisher={The Internet Society},
  doi={10.14722/ndss.2024.23131},
  url={https://www.ndss-symposium.org/ndss-paper/mock-optimizing-kernel-fuzzing-mutation-with-context-aware-dependency/}
}

@inproceedings{ji2026firmagent,
  title={FirmAgent: Leveraging Fuzzing to Assist LLM Agents with IoT Firmware Vulnerability Discovery},
  author={Ji, Jiangan and Zhang, Chao and Gan, Shuitao and Jian, Lin and Liu, Hangtian and Liu, Tieming and Zheng, Lei and Jia, Zhipeng},
  booktitle={Proceedings of the Network and Distributed System Security Symposium (NDSS 2026)},
  year={2026},
  publisher={The Internet Society},
  url={https://www.ndss-symposium.org/ndss-paper/firmagent-leveraging-fuzzing-to-assist-llm-agents-with-iot-firmware-vulnerability-discovery/}
}

@inproceedings{lee2026rtcon,
  title={RTCon: Context-Adaptive Function-Level Fuzzing for RTOS Kernels},
  author={Lee, Eunkyu and Park, Junyoung and Yun, Insu},
  booktitle={Proceedings of the Network and Distributed System Security Symposium (NDSS 2026)},
  year={2026},
  publisher={The Internet Society},
  url={https://www.ndss-symposium.org/ndss-paper/rtcon-context-adaptive-function-level-fuzzing-for-rtos-kernels/}
}

@misc{wang2026attentiondistance,
  title={Attention Distance: A Novel Metric for Directed Fuzzing with Large Language Models},
  author={Wang, Bin and Yang, Ao and Li, Kedan and Liu, Aofan and Li, Hui and Luo, Guibo and Huang, Weixiang and Zhuang, Yan},
  year={2025},
  doi={10.48550/arXiv.2512.19758},
  howpublished={\url{https://arxiv.org/abs/2512.19758}},
  note={arXiv:2512.19758 [cs.SE]. Accepted to ICSE 2026 Research Track}
}

@misc{shiraishi2026pilot,
  title={PILOT: Command-line Interface Fuzzing via Path-Guided, Iterative Large Language Model Prompting},
  author={Shiraishi, Momoko and Cao, Yinzhi and Shinagawa, Takahiro},
  year={2025},
  doi={10.48550/arXiv.2511.20555},
  howpublished={\url{https://arxiv.org/abs/2511.20555}},
  note={arXiv:2511.20555 [cs.CR]. Accepted at IEEE S\&P 2026}
}

@misc{tu2026cottontail,
  title={Cottontail: Large Language Model-Driven Concolic Execution for Highly Structured Test Input Generation},
  author={Tu, Haoxin and Lee, Seongmin and Li, Yuxian and Chen, Peng and Jiang, Lingxiao and B{\"o}hme, Marcel},
  year={2025},
  doi={10.48550/arXiv.2504.17542},
  howpublished={\url{https://arxiv.org/abs/2504.17542}},
  note={arXiv:2504.17542 [cs.SE]}
}

@misc{herter2025gptrace,
  title={GPTrace: Effective Crash Deduplication Using LLM Embeddings},
  author={Herter, Patrick and Ahlrichs, Vincent and A{\c{c}}ilan, Ridvan and Horsch, Julian},
  year={2025},
  doi={10.48550/arXiv.2512.01609},
  howpublished={\url{https://arxiv.org/abs/2512.01609}},
  note={arXiv:2512.01609 [cs.SE]. Accepted at ICSE 2026}
}

@misc{li2026portgpt,
  title={PORTGPT: Towards Automated Backporting Using Large Language Models},
  author={Li, Zhaoyang and Yu, Zheng and Song, Jingyi and Xu, Meng and Luo, Yuxuan and Mu, Dongliang},
  year={2025},
  doi={10.48550/arXiv.2510.22396},
  howpublished={\url{https://arxiv.org/abs/2510.22396}},
  note={arXiv:2510.22396 [cs.CR]. Accepted at IEEE S\&P 2026}
}

@Misc{openai2026gpt53codex,
  Title                    = {{Introducing GPT-5.3-Codex}},
  Author                   = {OpenAI},
  HowPublished             = {\url{https://openai.com/index/introducing-gpt-5-3-codex/}},
  year                     = {2026}
}

@Misc{openai2026gpt54,
  Title                    = {{Introducing GPT-5.4}},
  Author                   = {OpenAI},
  HowPublished             = {\url{https://openai.com/index/introducing-gpt-5-4/}},
  year                     = {2026}
}

@Misc{anthropic2026opus47,
  Title                    = {{Introducing Claude Opus 4.7}},
  Author                   = {Anthropic},
  HowPublished             = {\url{https://www.anthropic.com/news/claude-opus-4-7}},
  year                     = {2026}
}

@Misc{anthropic2026glasswing,
  Title                    = {{Project Glasswing: Securing Critical Software for the AI Era}},
  Author                   = {Anthropic},
  HowPublished             = {\url{https://www.anthropic.com/glasswing}},
  year                     = {2026}
}

@Misc{anthropic2026mythos,
  Title                    = {{Assessing Claude Mythos Preview's Cybersecurity Capabilities}},
  Author                   = {Anthropic Frontier Red Team},
  HowPublished             = {\url{https://red.anthropic.com/2026/mythos-preview/}},
  year                     = {2026}
}

@inproceedings{babic2019fudge,
  title={FUDGE: Fuzz Driver Generation at Scale},
  author={Babi{\'c}, Domagoj and Bucur, Stefan and Chen, Yaohui and Ivan{\v{c}}i{\'c}, Franjo and King, Tim and Kusano, Markus and Lemieux, Caroline and Szekeres, L{\'a}szl{\'o} and Wang, Wei},
  booktitle={Proceedings of the 2019 27th ACM Joint Meeting on European Software Engineering Conference and Symposium on the Foundations of Software Engineering},
  pages={975--985},
  year={2019},
  publisher={ACM},
  doi={10.1145/3338906.3340456},
  url={https://doi.org/10.1145/3338906.3340456}
}

@inproceedings{ispoglou2020fuzzgen,
  title={{FuzzGen}: Automatic Fuzzer Generation},
  author={Ispoglou, Kyriakos and Austin, Daniel and Mohan, Vishwath and Payer, Mathias},
  booktitle={29th USENIX Security Symposium (USENIX Security 20)},
  pages={2271--2287},
  year={2020},
  publisher={USENIX Association},
  url={https://www.usenix.org/conference/usenixsecurity20/presentation/ispoglou}
}

@inproceedings{chen2023hopper,
  title={Hopper: Interpretative Fuzzing for Libraries},
  author={Chen, Peng and Xie, Yuxuan and Lyu, Yunlong and Wang, Yuxiao and Chen, Hao},
  booktitle={Proceedings of the 2023 ACM SIGSAC Conference on Computer and Communications Security},
  pages={1600--1614},
  year={2023},
  publisher={ACM},
  doi={10.1145/3576915.3616610},
  url={https://doi.org/10.1145/3576915.3616610}
}

@Misc{ossfuzzgen,
  Title                    = {{OSS-Fuzz-Gen: A Framework for Fuzz Target Generation and Evaluation}},
  Author                   = {{Google}},
  HowPublished             = {\url{https://github.com/google/oss-fuzz-gen}},
  year                     = {2024}
}

@Misc{liu2023aipoweredfuzzing,
  Title                    = {{AI-Powered Fuzzing: Breaking the Bug Hunting Barrier}},
  Author                   = {Liu, Dongge and Metzman, Jonathan and Chang, Oliver and {Google Open Source Security Team}},
  HowPublished             = {\url{https://security.googleblog.com/2023/08/ai-powered-fuzzing-breaking-bug-hunting.html}},
  year                     = {2023}
}

@inproceedings{zhang2024llmfuzzdriver,
  title={How Effective Are They? Exploring Large Language Model Based Fuzz Driver Generation},
  author={Zhang, Cen and Zheng, Yaowen and Bai, Mingqiang and Li, Yeting and Ma, Wei and Xie, Xiaofei and Li, Yuekang and Sun, Limin and Liu, Yang},
  booktitle={Proceedings of the 33rd ACM SIGSOFT International Symposium on Software Testing and Analysis},
  pages={1223--1235},
  year={2024},
  publisher={ACM},
  doi={10.1145/3650212.3680355},
  url={https://doi.org/10.1145/3650212.3680355}
}

@inproceedings{lyu2024promptfuzz,
  title={Prompt Fuzzing for Fuzz Driver Generation},
  author={Lyu, Yunlong and Xie, Yuxuan and Chen, Peng and Chen, Hao},
  booktitle={Proceedings of the 2024 on ACM SIGSAC Conference on Computer and Communications Security},
  pages={3793--3807},
  year={2024},
  publisher={ACM},
  doi={10.1145/3658644.3670396},
  url={https://doi.org/10.1145/3658644.3670396}
}

@misc{xu2024ckgfuzzer,
  title={{CKGFuzzer}: {LLM}-Based Fuzz Driver Generation Enhanced By Code Knowledge Graph},
  author={Xu, Hanxiang and Ma, Wei and Zhou, Ting and Zhao, Yanjie and Chen, Kai and Hu, Qiang and Liu, Yang and Wang, Haoyu},
  year={2024},
  doi={10.48550/arXiv.2411.11532},
  howpublished={\url{https://arxiv.org/abs/2411.11532}},
  note={arXiv:2411.11532 [cs.SE]}
}

@inproceedings{liu2025promefuzz,
  title={{PromeFuzz}: A Knowledge-Driven Approach to Fuzzing Harness Generation with Large Language Models},
  author={Liu, Yuwei and Deng, Junquan and Jia, Xiangkun and Wang, Yanhao and Wang, Minghua and Huang, Lin and Wei, Tao and Su, Purui},
  booktitle={Proceedings of the 2025 ACM SIGSAC Conference on Computer and Communications Security},
  pages={1559--1573},
  year={2025},
  publisher={ACM},
  doi={10.1145/3719027.3765222},
  url={https://doi.org/10.1145/3719027.3765222}
}

@misc{yang2025harnessagent,
  title={{HarnessAgent}: Scaling Automatic Fuzzing Harness Construction with Tool-Augmented {LLM} Pipelines},
  author={Yang, Kang and Zhang, Yunhang and Li, Zichuan and Tao, GuanHong and Xu, Jun and Liao, XiaoJing},
  year={2025},
  doi={10.48550/arXiv.2512.03420},
  howpublished={\url{https://arxiv.org/abs/2512.03420}},
  note={arXiv:2512.03420 [cs.CR]}
}

@misc{ullah2025cvegenie,
  title={From {CVE} Entries to Verifiable Exploits: An Automated Multi-Agent Framework for Reproducing {CVEs}},
  author={Ullah, Saad and Balasubramanian, Praneeth and Guo, Wenbo and Burnett, Amanda and Pearce, Hammond and Kruegel, Christopher and Vigna, Giovanni and Stringhini, Gianluca},
  year={2025},
  doi={10.48550/arXiv.2509.01835},
  howpublished={\url{https://arxiv.org/abs/2509.01835}},
  note={arXiv:2509.01835 [cs.CR]}
}

@misc{wang2025cybergym,
  title={{CyberGym}: Evaluating {AI} Agents' Real-World Cybersecurity Capabilities at Scale},
  author={Wang, Zhun and Shi, Tianneng and He, Jingxuan and Cai, Matthew and Zhang, Jialin and Song, Dawn},
  year={2025},
  doi={10.48550/arXiv.2506.02548},
  howpublished={\url{https://arxiv.org/abs/2506.02548}},
  note={arXiv:2506.02548 [cs.CR]}
}

@inproceedings{wang2021syzvegas,
  title={$\{$SyzVegas$\}$: Beating kernel fuzzing odds with reinforcement learning},
  author={Wang, Daimeng and Zhang, Zheng and Zhang, Hang and Qian, Zhiyun and Krishnamurthy, Srikanth V and Abu-Ghazaleh, Nael},
  booktitle={30th USENIX Security Symposium (USENIX Security 21)},
  pages={2741--2758},
  year={2021}
}

@inproceedings{pham2020aflnet,
  title={Aflnet: A greybox fuzzer for network protocols},
  author={Pham, Van-Thuan and B{\"o}hme, Marcel and Roychoudhury, Abhik},
  booktitle={2020 IEEE 13th international conference on software testing, validation and verification (ICST)},
  pages={460--465},
  year={2020},
  organization={IEEE}
}

@inproceedings{atlidakis2019restler,
  title={Restler: Stateful rest api fuzzing},
  author={Atlidakis, Vaggelis and Godefroid, Patrice and Polishchuk, Marina},
  booktitle={2019 IEEE/ACM 41st International Conference on Software Engineering (ICSE)},
  pages={748--758},
  year={2019},
  organization={IEEE}
}

@inproceedings{xu2019janus,
  title={Fuzzing file systems via two-dimensional input space exploration},
  author={Xu, Wen and Moon, Hyungon and Kashyap, Sanidhya and Tseng, Po-Ning and Kim, Taesoo},
  booktitle={2019 IEEE Symposium on Security and Privacy (SP)},
  pages={818--834},
  year={2019},
  organization={IEEE}
}

@inproceedings{stephens2016driller,
  title={Driller: Augmenting fuzzing through selective symbolic execution.},
  author={Stephens, Nick and Grosen, John and Salls, Christopher and Dutcher, Andrew and Wang, Ruoyu and Corbetta, Jacopo and Shoshitaishvili, Yan and Kruegel, Christopher and Vigna, Giovanni},
  booktitle={NDSS},
  volume={16},
  number={2016},
  pages={1--16},
  year={2016}
}

@inproceedings{yun2018qsym,
  title={$\{$QSYM$\}$: A practical concolic execution engine tailored for hybrid fuzzing},
  author={Yun, Insu and Lee, Sangho and Xu, Meng and Jang, Yeongjin and Kim, Taesoo},
  booktitle={27th USENIX Security Symposium (USENIX Security 18)},
  pages={745--761},
  year={2018}
}

@inproceedings{kim2020hfl,
  title={HFL: Hybrid Fuzzing on the Linux Kernel.},
  author={Kim, Kyungtae and Jeong, Dae R and Kim, Chung Hwan and Jang, Yeongjin and Shin, Insik and Lee, Byoungyoung},
  booktitle={NDSS},
  year={2020}
}

@article{chipounov2011s2e,
  title={S2E: A platform for in-vivo multi-path analysis of software systems},
  author={Chipounov, Vitaly and Kuznetsov, Volodymyr and Candea, George},
  journal={Acm Sigplan Notices},
  volume={46},
  number={3},
  pages={265--278},
  year={2011},
  publisher={ACM New York, NY, USA}
}
\bibliographystyle{abbrv}
\clearpage
\appendix[Prompts Used in \texorpdfstring{\system}{\system}]

\label{app:prompts}

This appendix lists the task-specific user prompts used by \system. Each prompt
corresponds to one stage of the pipeline: synthesizing a fuzzing harness,
describing the harness interface in Syzkaller's syscall-description language,
and repairing integration-time compilation errors. Placeholders enclosed in
braces are instantiated with case-specific values during execution.

\newminted[prompttext]{text}{
  fontsize=\small,
  breaklines,
  breakautoindent=true,
  breakindent=1.5em,
  breaksymbolleft={},
  breaksymbolright={},
  frame=single,
  framesep=2mm,
  tabsize=2
}

\subsubsection{Harness Synthesis Prompt}

\noindent\textbf{Purpose.}
This prompt is used in the first stage. Given the patch,
the generation agent infers the trigger conditions and synthesizes a C++ fuzzing
harness. The harness fixes the high-level setup logic while exposing uncertain,
bug-critical syscall-related values as parameters of \texttt{entry()} so that
Syzkaller can later mutate them. This prompt corresponds the harness-generation described in \S\ref{sec:method-harness}.

\begin{prompttext}
You are given a patch that fixes a Linux kernel bug. Your task is to build a fuzzing harness that can later be fuzzed to trigger the bug fixed by the given patch.

First, analyze the trigger conditions by inspecting the patch message, code diff, and related source code. Then, based on the inferred trigger conditions, generate the fuzzing harness in C++.

The harness should contain:
1. an entry() function as the fuzzing entry point; and
2. a main() function.

If there are syscall-related variables whose values you are not sure how to set, you may expose them as parameters of entry(). In main(), assign random values to these parameters for testing, so that alternative assignments can be checked against the trigger conditions.

The patch is as follows:

```diff
{patch}
```

\end{prompttext}

\noindent\textbf{Variables used in this prompt.}
\begin{itemize}[leftmargin=*]
  \item \texttt{\{patch\}} contains the case-specific patch context, including the commit message and code diff used by the agent to infer trigger conditions.
  \item \texttt{entry()} is the generated harness entry point. Its parameters are the uncertain, bug-critical values that should remain fuzzable.
  \item \texttt{main()} is a local testing driver used by the agent to compile and sanity-check the harness before it is integrated into Syzkaller.
\end{itemize}

\subsubsection{Syscall-Description Generation Prompt}

\noindent\textbf{Purpose.}
After the harness is synthesized, this prompt asks the same generation agent to
describe the \texttt{entry()} interface in Syzkaller's syscall-description
language. This step maps each exposed \texttt{entry()} parameter to an appropriate
\texttt{syzlang} type, using Syzkaller documentation and existing descriptions as
references.

\begin{prompttext}
Your next task is to generate the syscall description for entry(). The entry() function will be treated as a Syzkaller pseudo-syscall.

Read /docs/syscall_syntax.md to understand the definition and syntax of Syzkaller syscall descriptions.

You may read files under /syzkaller_examples to inspect existing syscall descriptions.

In addition, the exposed arguments of entry() may already be described in existing syscall descriptions, either as standalone arguments or as fields of nested structures.

Finally, save the syscall description into a file named "syscall_description.txt" under the current directory.


\end{prompttext}

\noindent\textbf{Variables used in this prompt.}
\begin{itemize}[leftmargin=*]
  \item \texttt{entry()} is the harness interface generated in the previous stage and treated here as the pseudo-syscall to be described.
  \item The exposed arguments of \texttt{entry()} are the fuzzable variables selected during harness synthesis. This prompt asks the agent to assign each argument a compatible \texttt{syzlang} type.
  \item \texttt{/docs/syscall\_syntax.md} is the syntax reference for writing valid Syzkaller syscall descriptions.
  \item \texttt{/syzkaller\_examples} provides existing descriptions that can be used to infer suitable argument types and naming conventions.
  \item \texttt{syscall\_description.txt} is the required output file that stores the generated description.
\end{itemize}

\subsubsection{Compilation-Repair Prompt}

\noindent\textbf{Purpose.}
This prompt is used only if the injected pseudo-syscall or syscall description
causes Syzkaller to fail during generation or compilation. It gives the repair
agent the compiler log and the exact editable line ranges, and it constrains the
agent to fix only the newly injected artifacts rather than modifying unrelated
Syzkaller code.

\begin{prompttext}
You are fixing compilation errors in Syzkaller after pseudo-syscall injection.

## Compile Error
{error_log}

## Injected Code Locations
### 1. Pseudo Syscall Code
- File: executor/common_linux.h
- Lines: {pseudo_syscall_start} through {pseudo_syscall_end}
- This code is between the `// AUTO-INJECTED PSEUDO SYSCALL` and `// END AUTO-INJECTED` markers

### 2. Syscall Description
- File: sys/linux/dev_vtpm.txt
- Lines: {syscall_desc_start} through {syscall_desc_end}
- This is the entire file content created by the pipeline

## Instructions
1. **Analyze the error log** to determine which file caused the error:
   - If the error mentions `executor/common_linux.h`, the problem is in the pseudo-syscall code
   - If the error mentions `sys/linux/dev_vtpm.txt` or occurs during `make generate`, the problem is in the syscall description
   - Errors may be caused by both files
   - Even if the error mentions OTHER files, the root cause is ALWAYS in our injected code

2. Read the relevant file(s) based on your analysis
3. Fix the compile errors in the INJECTED CODE ONLY
4. Run `make` to verify the fix
5. If compilation still fails, repeat from step 1

## CRITICAL RULES
- You can ONLY modify these two files:
  - executor/common_linux.h (only lines {pseudo_syscall_start} through {pseudo_syscall_end})
  - sys/linux/dev_vtpm.txt (only lines {syscall_desc_start} through {syscall_desc_end})
- DO NOT modify any other files, even if errors mention them
- Errors in other Syzkaller files are CAUSED BY our injected code; fix the cause, not the symptom
- Assume every error is caused by the pseudo-syscall code, the syscall description, or both

\end{prompttext}

\noindent\textbf{Variables used in this prompt.}
\begin{itemize}[leftmargin=*]
  \item \texttt{\{error\_log\}} is the Syzkaller generation or compilation error log observed after injecting the new artifacts.
  \item \texttt{\{pseudo\_syscall\_start\}} and \texttt{\{pseudo\_syscall\_end\}} delimit the editable region in \texttt{executor/common\_linux.h} that contains the injected pseudo-syscall implementation.
  \item \texttt{\{syscall\_desc\_start\}} and \texttt{\{syscall\_desc\_end\}} delimit the editable region in \texttt{sys/linux/dev\_vtpm.txt} that contains the generated syscall description.
  \item \texttt{executor/common\_linux.h} is the Syzkaller executor file into which the pseudo-syscall body is injected.
  \item \texttt{sys/linux/dev\_vtpm.txt} is simply the existing Syzkaller syscall-description file that we chose as the insertion point for the generated pseudo-syscall description. This choice is arbitrary rather than semantically meaningful: the generated description could equally be injected into any other existing syscall-description file.
\end{itemize}

\end{document}